\documentclass{article}

\usepackage[english]{babel}

\usepackage[letterpaper,top=2cm,bottom=2cm,left=3cm,right=3cm,marginparwidth=1.75cm]{geometry}

\usepackage{amsmath}
\usepackage{tikz}
\usepackage{color}
\usepackage{etoolbox}
\usepackage{graphicx}
\usepackage{blindtext}
\usepackage{subfiles}          
\usepackage{authblk}           
\usepackage{caption}           
\usepackage{subcaption}        
\usepackage{indentfirst}       
\usepackage[backend=biber,style=phys,sorting=none,max names=4]{biblatex}          
\usepackage{hyperref}
\hypersetup{colorlinks=true, allcolors=black}

\definecolor{mydarkblue}{rgb}{0.2422, 0.1504, 0.6603}
\definecolor{mylightblue}{rgb}{0.1540, 0.5902, 0.9218}
\definecolor{mygreen}{rgb}{0.5044, 0.7993, 0.3480}
\definecolor{myyellow}{rgb}{0.9769, 0.9839, 0.0805}
\newrobustcmd*{\myfilledsquare}[1]{\tikz{\filldraw[draw=black,fill=#1] (0,0)
rectangle (0.15cm,0.15cm);}}
\newrobustcmd*{\myemptysquare}[1]{\tikz{\draw[draw=#1,thick] (0,0)
rectangle (0.15cm,0.15cm);}}
\newrobustcmd*{\myfilledcircle}[1]{\tikz{\filldraw[draw=black,fill=#1] (0,0)
circle [radius=0.1cm];}}
\newrobustcmd*{\myemptycircle}[1]{\tikz{\draw[draw=#1, very thick] (0,0)
circle [radius=0.1cm];}}
\newrobustcmd*{\myfilleduptriangle}[1]{\tikz{\filldraw[draw=black,fill=#1] (0,0) --
(0.2cm,0) -- (0.1cm,0.2cm) -- (0,0);}}
\newrobustcmd*{\myemptyuptriangle}[1]{\tikz{\draw[draw=#1, very thick] (0,0) --
(0.2cm,0) -- (0.1cm,0.2cm) -- (0,0);}}
\newrobustcmd*{\myfilleddowntriangle}[1]{\tikz{\filldraw[draw=black,fill=#1] (0,0) --
(-0.2cm,0) -- (-0.1cm,-0.2cm) -- (0,0);}}
\newrobustcmd*{\myemptydowntriangle}[1]{\tikz{\draw[draw=#1, very thick] (0,0) --
(-0.2cm,0) -- (-0.1cm,-0.2cm) -- (0,0);}}
\newrobustcmd*{\myfilledlefttriangle}[1]{\tikz{\filldraw[draw=black,fill=#1] (-0.1cm,0) --
(0.1cm,0.1cm) -- (0.1cm,-0.1cm) -- (-0.1cm,0);}}
\newrobustcmd*{\myemptylefttriangle}[1]{\tikz{\draw[draw=#1,very thick] (-0.1cm,0) --
(0.1cm,0.1cm) -- (0.1cm,-0.1cm) -- (-0.1cm,0);}}

\newrobustcmd*{\myfilledrighttriangle}[1]{\tikz{\filldraw[draw=black,fill=#1] (0.1cm,0) --
(-0.1cm,0.1cm) -- (-0.1cm,-0.1cm) -- (0.1cm,0);}}
\newrobustcmd*{\myemptyrighttriangle}[1]{\tikz{\draw[draw=#1, very thick] (0.1cm,0) --
(-0.1cm,0.1cm) -- (-0.1cm,-0.1cm) -- (0.1cm,0);}}
\newrobustcmd*{\myemptydiamond}[1]{\tikz{\draw[draw=#1, very thick] (0,-0.15cm) --
(-0.1cm,0) -- (0,0.15cm) -- (0.1cm,0) -- (0,-0.15cm);}}

\title{On the interaction between the island divertor heat fluxes, the scrape-off layer radial electric field and the edge turbulence in Wendelstein 7-X plasmas}
\author[1]{E. Maragkoudakis}
\author[1]{D. Carralero}
\author[1]{T. Estrada}
\author[2]{T. Windisch}
\author[2]{Y. Gao}
\author[2]{C. Killer}
\author[2]{M. Jakubowski}
\author[2]{A. Puig Sitjes}
\author[3]{F. Pisano}
\author[4]{H. Sándor}
\author[4]{M. Vecsei}
\author[4]{S. Zoletnik}
\author[1]{A. Cappa}
\author[ ]{the Wendelstein 7-X team}

\affil[1]{Laboratorio Nacional de Fusión, CIEMAT, 28040 Madrid, Spain.}
\affil[2]{Max-Planck-Insitut für Plasmaphysik, D-17491 Greifswald, Germany}
\affil[3]{Department of Electrical and Electronic Engineering, University of Cagliari, Italy}
\affil[4]{Fusion Plasma Physics Department, Centre for Energy Research, Budapest, Hungary}

\date{}

\begin{document}

\maketitle

\begin{abstract}

The formation of the radial electric field, $E_{\rm r}$ in the SOL has been experimentally studied for attached divertor conditions in stellarator W7-X. The main objective of this study is to test the validity in a complex 3D island divertor of simple models, typically developed in tokamaks, relating $E_{\rm r}$ to the sheath potential drop gradient at the target. Additionally, we investigate the effect of the edge $E_{\rm r}$ shear on the reduction of density fluctuation amplitude, a well-established phenomenon according to the existing bibliography. The main diagnostic for measurements in the SOL is a V-band Doppler reflectometer that can provide the measurement of the $E_{\rm r}$ and density fluctuations with good spatial resolution. Three-dimensional measurements of divertor parameters has been carried out using infrared cameras, with $\lambda_q$ resulting a suitable proxy for the model-relevant $\lambda_T$. In the investigated attached regimes, it is shown for the first time that the formation of the $E_{\rm r}$ in the SOL depends on parameters at the divertor, following a $E_{\rm r} \propto T_e/\lambda_q$ qualitatively similar to that found in a tokamak. Then, from the analyzed plasmas, the observed $E_{\rm r}$ shear at the edge is linked to a moderate local reduction of the amplitude of density fluctuations.

\end{abstract}

\section{Introduction}

The region surrounding the nested magnetic surfaces of a magnetic confinement device is truly important for the realistic development of a commercial fusion reactor, as it determines how the particle and energy fluxes are deposited onto the various surfaces facing the plasma. This region is typically referred as the Scrape-Off Layer (SOL) and can feature various degrees of complexity: from the relatively simple and symmetric SOL of a divertor tokamak, to the asymmetric geometry of a stellarator equipped with an island divertor involving an island chain with secondary regions around the O-points. A simple and general first approach is that SOL flows follow the open field lines onto the material surfaces that intersect them. However, this parallel transport must be compounded with substantial cross-field flows, typically referred as "drifts". Among them, perhaps the most important for SOL flows is the $v_{E \times B}$ drift, caused by local gradients in electrostatic potential, $\phi$. These drifts may strongly influence the position and intensity of heat loads onto divertor targets \cite{Goldston2012} and have been proposed in the literature as the underlying mechanism for the observed asymmetries between the high field side and low field side divertors in tokamaks (see e.g. \cite{Rozhansky2012} and references therein). Because of this, a good understanding of the process of electric field formation in the SOL is of paramount importance for the optimized helias stellarator Wendelstein 7-X (W7-X) \cite{Grieger1992,Bosch2013,Klinger2017}. This device, the largest of its kind currently in operation, achieved its first plasmas in 2016 \cite{Klinger2017} and has among its main objectives to test the island divertor as a valid plasma-wall interaction solution for a helias reactor \cite{Wolf2019}. Besides its importance for SOL flows, the electric field in the SOL may also play a relevant role on the confinement and performance of the plasma in the region of closed flux surfaces, also referred as region of confined plasma: Typical W7-X plasmas feature a sheared $E_{\rm r}$ in the edge, as the negative ion root radial electric field in the confined region switches to the positive values usually found in the SOL. This creates a strong shear in the perpendicular $v_{E \times B}$ velocity, which has been predicted to suppress turbulence and lead to the formation transport barriers \cite{Biglari1990}. The structure of the radial electric field around the LCFS has been found to play a key role in the transition to the improved confinement H-mode in tokamaks \cite{Taylor1989, Moyer2001, Viezzer2014} and stellarators \cite{Brakel1997, Happel2011} (although in some cases this $E_r$ has been found to be the result rather than the cause of the transition \cite{Andrew2008, Estrada2009}). The majority of these studies have focused mainly on the negative branch of the radial electric field found in the confined region. However, a number of recent works suggest that the $E_{\rm r}$ value at the SOL might have also a direct impact on global confinement, either by setting the flow boundary condition at the SOL \cite{LaBombard2005} or by enhancing the velocity shear and the suppression of edge turbulence \cite{Chankin2019}. In the latter case, a feedback mechanism is proposed which would link pedestal, SOL and target: as turbulence in the edge is reduced by the $E_{\rm r}$ shear, so is perpendicular anomalous particle and heat transport at the SOL, which would reinforce the velocity shear initially suppressing edge turbulence. Although no systematic study has been carried out on the effect of $E_{\rm r}$ shear on the turbulent transport in the edge of W7-X, there are some indications that it might be leading to the suppression of fluctuations \cite{KrAmer-Flecken2020}. Also an investigation of the experimental scenarios of the last W7-X campaign indicated that the edge shear of $E_{\rm r}$ is strongly influenced by the value of $E_r$ at the SOL \cite{Carralero2020}. Therefore, investigating the link between the SOL $E_r$ formation, targets and turbulence suppression is certainly relevant in this device.\\

The formation of SOL electric fields in tokamaks has been extensively studied in recent years and many works on this subject can be found in literature. As a result, some simplified models \cite{Stangeby1996} have been found to reasonably describe mid-plane potential profiles and electric fields. In general, upstream potential ($\phi_{u}$) at a given radial position $r$ can be defined as 

\begin{equation}
    \phi_{u}(r,s) = \phi_{s}(r) + \int_{0}^{s} E_{\parallel}(r,s) \, ds
    \label{eq:potentialGeneral}
\end{equation}
where $s$ is the coordinate along the field line, $s=0$ corresponds to the sheath entrance and $\phi_{s}(r)$ is the potential at the sheath entrance, which for a material with conductive surface can be usually approximated taking the potential drop at the sheath, $\phi_{s}(r) \simeq 3 T_{e}(r) / e$. Regarding the parallel electric field, it can be obtained from the momentum equation for electrons along the $\textbf{B}$ direction,

\begin{equation}
    E_{\parallel} = \frac{j_{\parallel}}{\sigma_{\parallel}} + \frac{1}{en_{e}} \frac{\partial p_{\rm e}}{\partial s} + 0,71 \frac{1}{e} \frac{\partial T_{\rm e}}{\partial s}
    \label{eq:parallelE}
\end{equation}
where $j_{\parallel}$ and $\sigma_{\parallel}$ are the parallel current and electric conductivity and $p_{\rm e}$ is the electron pressure. 

The simplest situation is found for low density, attached divertor conditions (the so called sheath limited regime). In this case, resistivity is low over the SOL ($j_{\parallel} / \sigma_{\parallel} \simeq 0$). Pressure can be considered roughly constant along the field line ($p_{\rm e} \simeq \rm{const}$). As well, $T_{\rm e}$ gradients parallel to the magnetic field tend to be small across the SOL. Thus, the rhs of equation \ref{eq:potentialGeneral} is dominated by the first term and $\phi_{u}(r) \simeq \phi_{s}(r)$. This means that, for low densities, $v_{E \times B}$ is mostly poloidal as it is dominated by the radial component of the electric field ($E_{r,u}$) which in turn, is determined by the gradient of the electron temperature in front of the target. Indeed, assuming an exponential decay of the temperature at the sheath entrance $T_{e,s} \simeq T_{e,0} \rm{exp}(-r/\lambda_{T})$, the radial electric field can be expressed as:

\begin{equation}
    E_{\rm{r},u} \equiv - \frac{\partial \phi_u}{\partial r} \simeq - \frac{3}{e} \frac{T_{e,s}}{\lambda_{T}}
    \label{eq:finalEr}
\end{equation}

As the density at the separatrix is increased, this situation becomes more complex: first, in the conduction dominated regime significant gradients of $T_e$ appear along the field line and as the recycling in the target increases, charge exchange and other atomic processes begin to drain momentum from the plasma flow. However, it is when detachment is significant that the simple low density picture loses its validity: as the heat fluxes onto the target are greatly reduced around the strike point, electron temperatures at the target drop and their profile becomes substantially flatter. Eventually, collisionality is increased to the point in which the first term in the rhs of equation \ref{eq:parallelE} becomes dominant and the parallel electric field is raised by parallel currents. In this new situation, potential differences appear between poloidally adjacent flux tubes, leading to poloidal electric fields and strong radial drifts. As a result, this dominant electric field is related to thermoelectric or Pfirsch-Schlütter return currents and no longer bound to the $T_e$ gradients at the target.\\

This physical picture is generally consistent with recent experimental results from tokamak plasmas: on the one hand, the simple relation between upstream radial electric fields at the target expressed in equation \ref{eq:finalEr} has been verified for attached plasmas in DIII-D \cite{Jaervinen2018}, JET \cite{Silva2021} and AUG \cite{BridaPSI2022}. Also, experiments carried out in detached plasmas have shown how $\phi_{u}$ is dominated by the parallel current contribution of the $E_{\parallel}$ thus making the SOL drifts to be dominated by the $E_{\theta} \times B$ term, $\theta$ being the poloidal direction. Examples can be found in DIII-D \cite{Jaervinen2018} and TCV \cite{Wensing2020} tokamaks. This kind of investigation is less frequently found in stellarators, where the precise mechanism leading to the formation of electric fields in the SOL is not as clear. The main reason for this is the fully three-dimensional topology of their SOL, which substantially increases the complexity of both experiments and theoretical models. Nevertheless, some investigations have been carried out in stellarators featuring island divertors which indicate that the underlying mechanisms behind SOL could be qualitatively similar to those in tokamaks. In \cite{Grigull2003}, flows onto the targets were investigated in divertor island configurations at Wendelstein 7-AS. For densities below detachment, strong up-down asymmetric shifts of the ion saturation current and $T_{\rm{e},s}$ peaks are observed which nearly reverse when magnetic field direction is inverted. This was interpreted as a strong indication of $v_{E \times B}$ drifts dominating  cross-field SOL flows, in good agreement with EMC3-EIRENE simulations, which found the electric field to be dominated by the radial component associated to the sheath potential gradient \cite{Feng2003, Feng1999, Feng1999a}. Interestingly, the impact of drifts is found to be stronger than in a tokamak divertor with the same plasma parameters due to the longer connection lengths. More recently, similar results were obtained in reversed field experiments carried out in W7-X \cite{Hammond2019}. In those experiments, for low density operation, strong up-down asymmetries were found on the radial positions of the strike lines and heat load distributions onto the targets, which could be attributed to poloidal drifts originating from dominant $E_{\rm r}$ fields. In good agreement with this, the asymmetries were again reversed when the direction of the magnetic field was changed. However, when the same measurements were carried for higher densities, up-down asymmetries became much weaker and a general inwards shift of the heat and particle flows was observed. This would be the result of a more complex drift structure, in which $E_{\rm r}$ is weakened by the shallower gradients of $T_{\rm e}$ at the targets and substantial parallel gradients of temperature, giving rise to non-negligible $E_{\parallel}$.\\

In this work, we set out to investigate the mechanism giving rise to the radial electric field at the SOL for W7-X and its link to the suppression of edge turbulence. For this, we carry out upstream radial electric field measurements at the vicinity of the separatrix using a Doppler reflectometer and use a field line tracing code to follow the field lines projecting the measurement locations onto the corresponding target. Since this results in a complex 2D projection, we then resort to infrared imaging in order to obtain a heat flux profile corresponding to the line of sight of the reflectometer. Applying this analysis to a database of discharges covering a range of plasmas in the attached divertor regime for a given magnetic configuration, we are able to find for the first time in a stellarator with island divertor configuration, direct evidence of the upstream-downstream interaction mechanism described in equation \ref{eq:finalEr}. Besides, using a wider database of Doppler reflectometry measurements, we show how the observed variation in $E_r$ shear moderately decreases the amplitude of density fluctuations across the separatrix. The remainder of this paper is organized as follows: The experimental set up along with elements of the analysis method will be described in section \ref{section:experimental_setup}. All results will be presented in section \ref{section:results}. Finally, the summary and discussion are given in section \ref{section:conclusion}.

\label{section:introduction}

\section{Experimental setup}

The experiments reported in this article were performed at the superconducting optimized helias stellarator W7-X with major radius $R=5.5\rm{m}$, minor radius $a = 0.5 \rm{m}$ and magnetic field on axis $B_0 = 2.5 \rm{T}$ \cite{Klinger2017}. W7-X uses an ECH system of 10 long-pulsed gyrotrons (each delivering $0.8 \rm{MW}$) \cite{Erckmann2017}. In each of its 5 periods, W7-X uses a set of 10 non-planar ("NPC") and 4 planar ("PC") superconductive coils in order to produce the confining magnetic field. Along one period, the plasma cross-section changes from bean shape to triangular shape. W7-X can employ a number of configurations \cite{Geiger2014}. Under a number of these configurations, resonant magnetic islands form at the plasma boundary which are intersected by the divertor plates creating the island-divertor \cite{Renner2002,Klinger2019}. This set of resonant magnetic islands surrounding the last closed flux surface (LCFS) is created by the non-planar coils while extra control over the magnetic topology of the islands is achieved through the use of another set of copper coils, the control coils ("CC") \cite{Feng2006, Feng2021}. As will be explained later, this work concerns plasmas in the standard configuration of W7-X. For the standard configuration, a 5/5 island chain\footnote{Meaning 5 islands that wind around the stellarator torus 5 times.} is created around the LCFS. Those islands are intersected periodically by the divertor. In each of the 5 periods of W7-X, there are 2 divertor modules, on the upper ("u") and the lower ("l") part. Each divertor module has two plates, the vertical ("v") and the horizontal ("h"). Thus, by referring to s1lh, we refer to module 1, lower horizontal target plate. For an in depth description of the island divertor concept, the reader is referred to \cite{Pedersen2018}. \\

\begin{figure}[!ht]
\begin{subfigure}{0.32\textwidth}
  \centering
  \includegraphics[width=\textwidth]{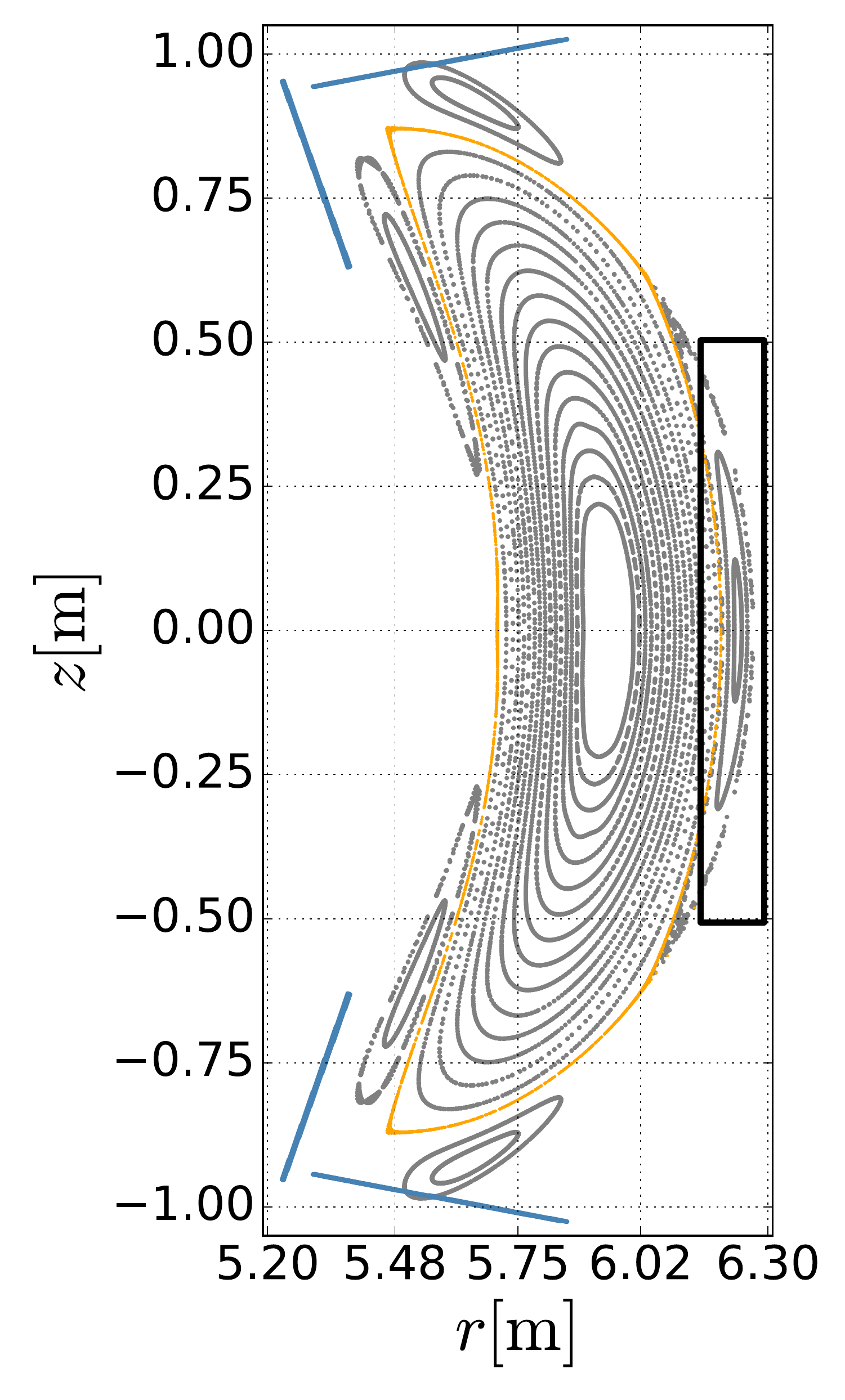}
  \caption{}
  \label{subfigure:methodPoincare}
\end{subfigure}
\begin{subfigure}{0.32\textwidth}
  \centering
  \includegraphics[width=\textwidth]{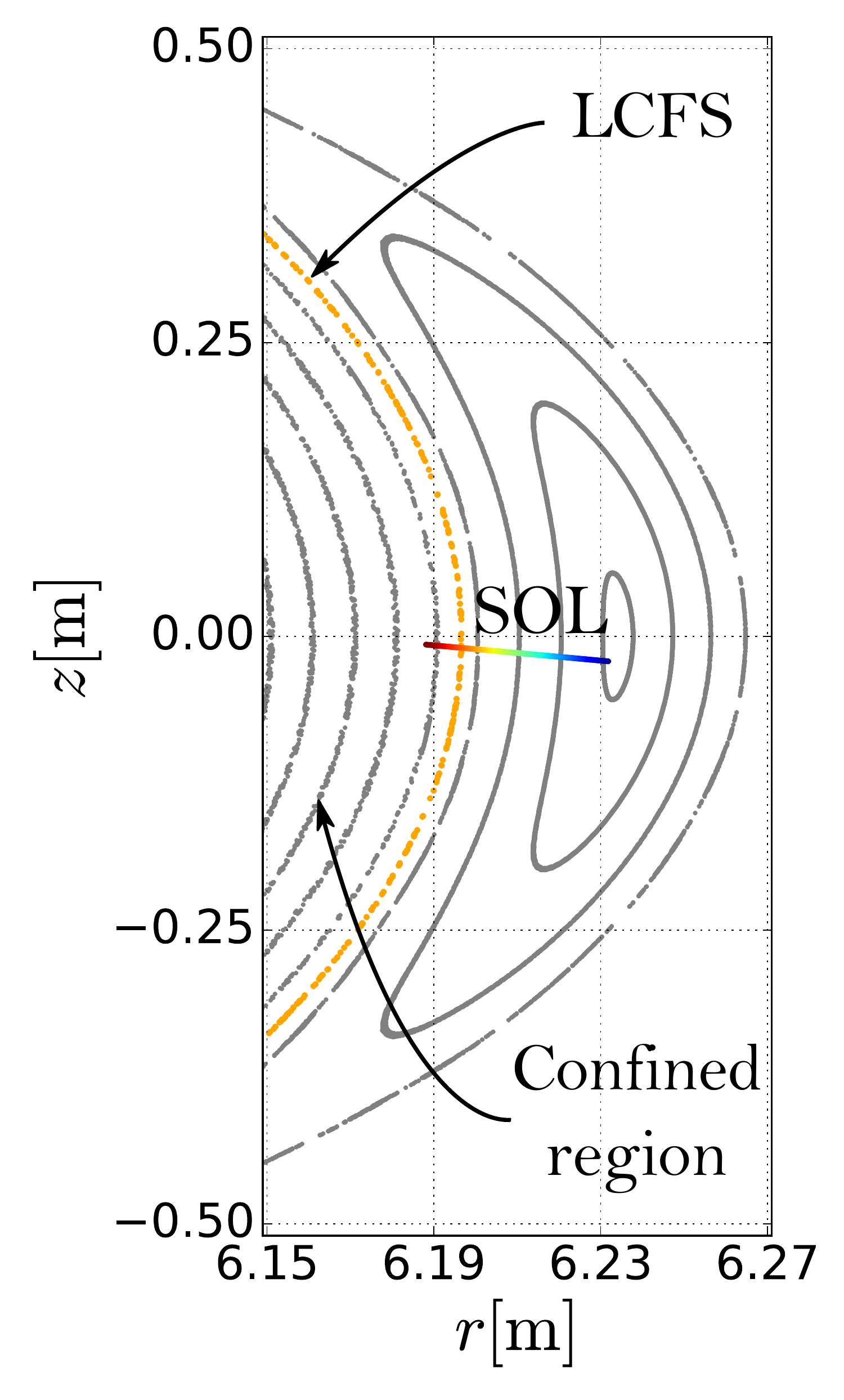}
  \caption{}
  \label{subfigure:methodPoincareZoom}
\end{subfigure}
\begin{subfigure}{0.32\textwidth}
  \centering
  \includegraphics[width=\textwidth]{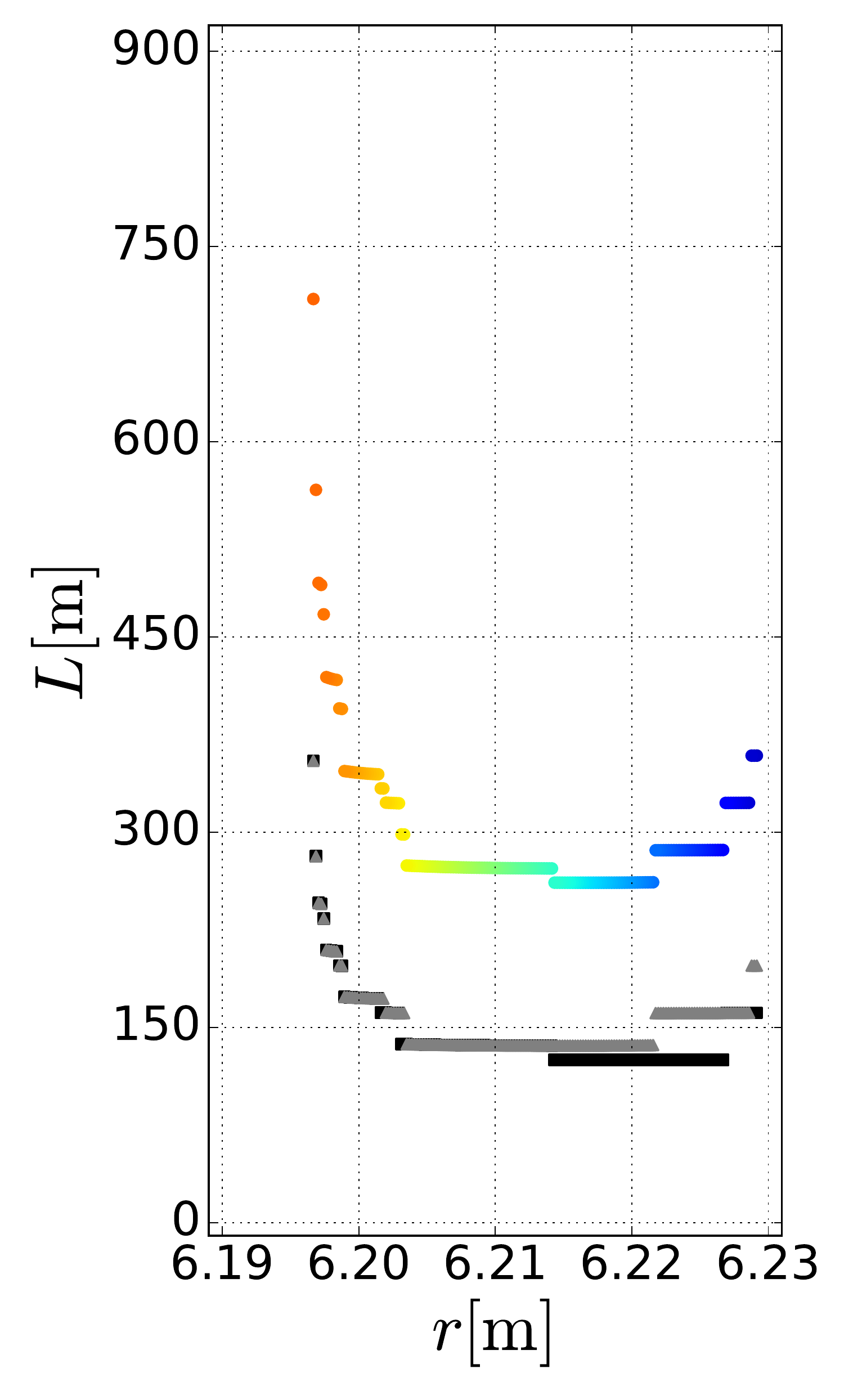}
  \caption{}
  \label{subfigure:connectionLength}
\end{subfigure}
\caption{(a) Poincaré plot of the standard magnetic configuration of W7-X for toroidal angle of $72^{\circ}$ (W7-X program 20180905.017, $t = 4,750 \rm ms$ to $5,000 \rm ms$). The divertor target plates are depicted as blue lines, while the LCFS is depicted with orange color. The region inside the rectangle, also in (b), with the island crossing the mid-plane ($z = 0$), is the region in which measurements are discussed later. The colored line represents the line of sight of the Doppler reflectometer in the SOL. In (c), the connection length calculated in the measurement area.}
\label{figure:methodPoincare}
\end{figure}

The Poincaré plot of W7-X standard magnetic configuration at toroidal angle equal to $72^{\circ}$ is obtained using a field line tracing (FLT) code \cite{Bozhenkov2013}, which calculates the magnetic field using the Biot-Savart law using as input coil currents. The result is shown in plot \ref{subfigure:methodPoincare}: The 5/5 island chain is depicted outside the LCFS, which is indicated with orange color. For this toroidal angle, the magnetic field cross-section is bean-shaped and the divertor targets, depicted as blue diagonal lines, intersect the top and bottom islands. The island crossing the mid-plane ($z=0$, in the rectangle) is drawn in more detail in plot \ref{subfigure:methodPoincareZoom}. The colored line represents the line of sight of the Doppler reflectometer, one of the main diagnostics used for this work that will be discussed later. As mentioned in the introduction, the higher the connection length of the island-divertor SOL field lines, the stronger the impact of drifts in the SOL transport. The field lines of the W7-X SOL have connection lengths of a few hundred meters \cite{Killer2019}, as can also be seen in plot \ref{subfigure:connectionLength}, where the connection length to the target or other physical components ($L$) was calculated along the line of sight of the Doppler reflectometer. As well in the same plot, the SOL is defined as the region with values of $L < 350 \rm m$. For lower or higher values of the toroidal radial coordinate $r$, we encounter field lines of infinite $L$ which correspond to the confined region inside of the LCFS or to the region in the vicinity of the O-point of the island, the referred secondary confined regions. \\

In order to investigate the topics presented in the introduction, we need to measure the $E_{\rm r}$ in the SOL, the density fluctuation amplitude and parameters over the divertor surface of W7-X. For the measurement of the radial electric field and the density fluctuation amplitude, we use the V-band monostatic frequency hopping Doppler reflectometer (DR) \cite{windisch201542nd, windisch2019w}. The DR is installed at the AEA-21 port (toroidal angle $72^{\circ}$). Its line of sight (depicted as colored line) is crossing the island of figure \ref{subfigure:methodPoincareZoom}. A DR differs from conventional reflectometry in the finite angle between finite angle between the probing wave and the cut-off layer normal, which is set in order to separate the Bragg back-scattered wave from the reflected wave at cut-off layer. By measuring the Doppler shifted back-scattered wave, the plasma turbulence and its perpendicular rotation velocity can be obtained. The latter is a composition of both the plasma $E\times B$ velocity and the intrinsic phase velocity of the density turbulence:  $v_{\perp} = v_{E\times B} + v_{\rm ph}$. In cases in which the condition $v_{E\times B} \gg v_{\rm ph}$ holds, the radial electric field can be obtained directly from the perpendicular rotation velocity:  $E_{\rm r} = B \cdot  v_{\perp} = B \cdot 2\pi f_D/k_\perp $, where $f_D$ is the frequency of the Doppler peak and $B$ is the magnetic field. Experiments performed in several devices have demonstrated that this assumption is, in general, valid \cite{Hirsch2001, Conway2004,Manz2018, windisch2019w}, and only in high collisionality plasmas, the contribution of $v_{\rm ph}$ to $v_{\perp}$ may become relevant \cite{Estrada2019, Vermare2011}. Besides, previous results obtained in W7-X show a good agreement between the experimental $E_{\rm r}$ profiles and those obtained from neoclassical predictions \cite{Carralero2020},\cite{Estrada2021}. Thus, it is sensible to assume that this condition holds also in the present experiments. Regarding the density fluctuations, the power of the back-scattered DR signal, $S$, is the relevant quantity proportional to  $\delta n_{e}^2$ and is given by $S = A_D \cdot \Delta f_D$, where $A_D$ and $\Delta f_D$ are the height and width of the Doppler peak. It has to be noted that, in general, a microwave generator working with variable frequency produces a different power output at each frequency. Besides, the transmitted power through the transmission line may also depend on the frequency. Therefore, a power calibration of the Doppler reflectometer is indispensable for a proper comparison of the fluctuations measured at different frequencies.\\

\begin{figure}[!ht]
\begin{subfigure}{0.49\textwidth}
  \centering
  \includegraphics[width=\textwidth]{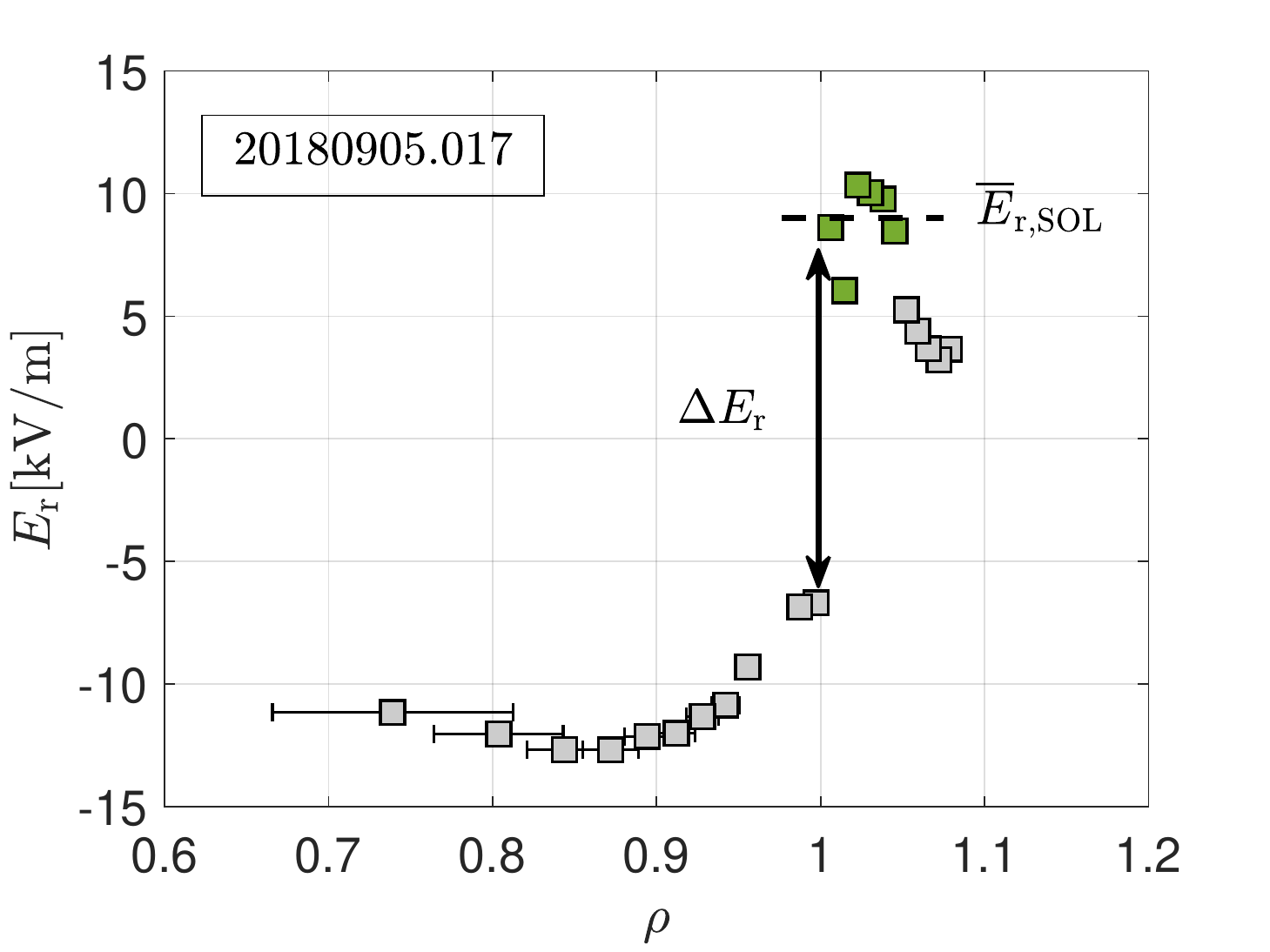}
  \caption{}
  \label{subfigure:180905017_profileRadialE}
\end{subfigure}
\begin{subfigure}{0.49\textwidth}
  \centering
  \includegraphics[width=\textwidth]{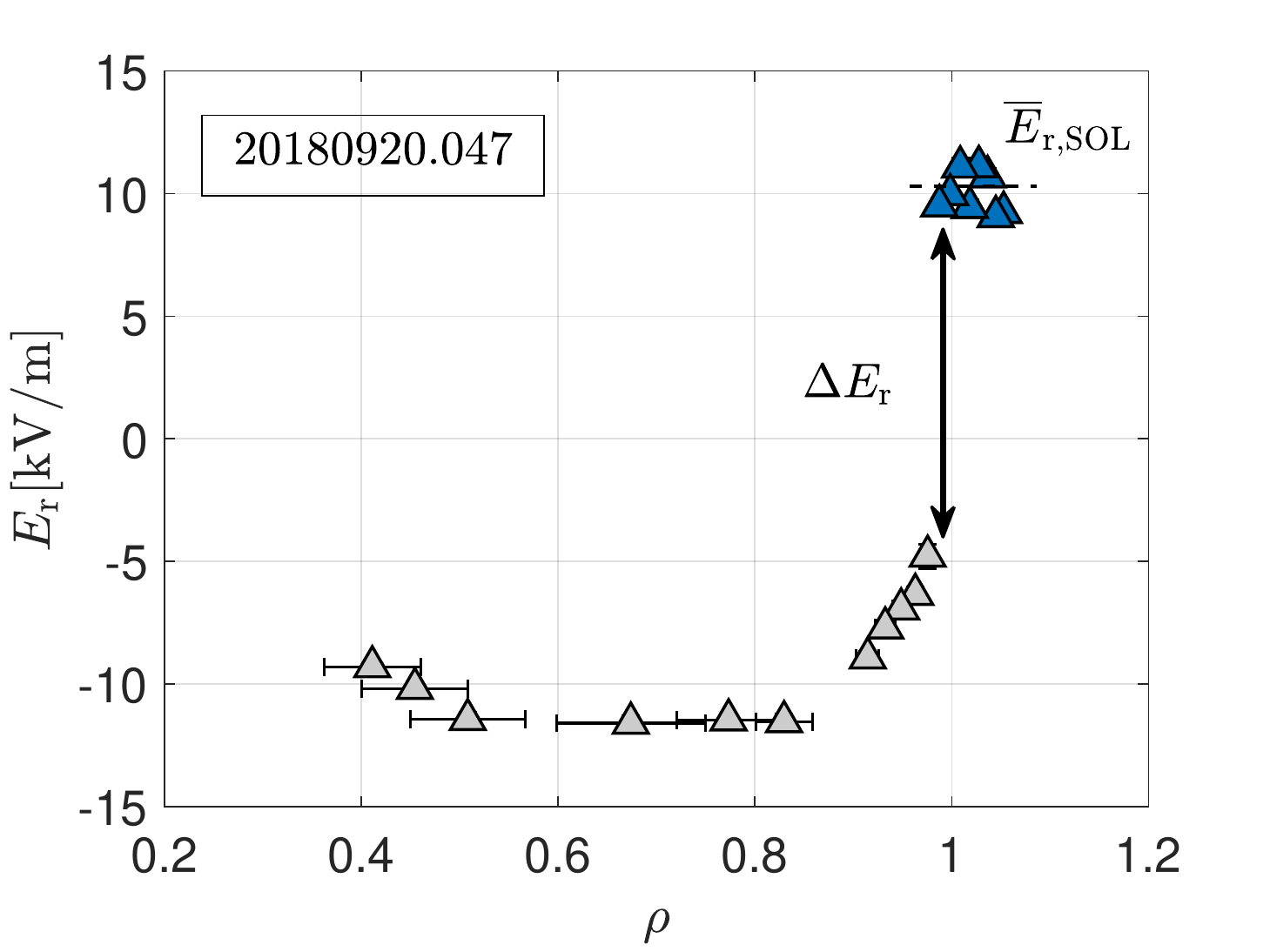}
  \caption{}
  \label{subfigure:180920047_profileRadialE}
\end{subfigure}
\caption{In figure \ref{subfigure:180905017_profileRadialE} and \ref{subfigure:180920047_profileRadialE}, the radial electric field profile as given by the DR for program 20180905.017 ($t = 4,750\rm ms$ to $t = 5,000 \rm ms$) and 20180920.047 ($t = 7,000\rm ms$ to $t = 7,250 \rm ms$).}
\label{figure:betterProfileEr}
\end{figure}

For this work, the DR was programmed to use frequencies from $50\rm{GHz}$ - $74\rm{GHz}$ in ordinary mode polarization (O-mode) and with a fixed injection angle of $18^{\circ}$. A full frequency scan was set to last $250$\rm{ms} with frequency steps of $1\rm{GHz}$. For the calculation of the position of the cut-off and the $k_{\perp}$ of the cut-off of the DR wave, we use the ray tracing code Travis \cite{Marushchenko2014} that takes as input the density profile and magnetic configuration. For this set-up, the expected values of $k_{\perp}$ are between $7-10$ cm$^{-1}$. Density measurements were provided by Thomson scattering (TS) \cite{Pasch2016} and the mapping of those measurements along the flux surface coordinate $\rho$ was calculated using VMEC equilibria \cite{Hirshman1983}. For the estimation of uncertainties on the cut-off position and $k_{\perp}$, the reflection of a bundle of rays is considered in Travis, with a width equal to the $1/e$ amplitude of the DR probing beam. Two typical examples of $E_{\rm r}$ profiles are shown in plot \ref{subfigure:180905017_profileRadialE} for program 20180905.017 ($t = 4,750 \rm ms$ to $t = 5,000 \rm ms$) and plot \ref{subfigure:180920047_profileRadialE} for program 20180920.047 ($t = 7,000\rm ms$ to $t = 7,250 \rm ms$). These particular discharges are presented and used to show paradigmatic data from our database. A detailed explanation about the calculation of the vertical and horizontal uncertainty bars in the $E_{\rm r}$ profiles can be found in \cite{Carralero2020}.\\ 

A few remarks on the DR data are in order: First, the DR can provide measurements for the region of open and closed field lines. It was seen that there is a sufficient number of DR measurements in the SOL for plasmas with $\overline{n}_{\rm e} > 4.5 \times 10^{19} \rm{m^{-3}}$. The transition from the SOL to the confined region inside the LCFS is seen in the DR signal as a sign reversal of the Doppler shift that regularly appears around the LCFS when the negative ion root $E_{\rm r}$ inside the LCFS \cite{Pablant2018, Windisch2017}, changes to the positive $E_{\rm r}$ at the SOL, giving the rise to the observed velocity shear of the $E_{\rm r}$ at the edge. The variation of the $E_{\rm r}$ in this part of the plasma occurs in radial scales smaller than a) the separation of the TS channels in this region \cite{Bozhenkov2017} and b) the radial resolution of the DR \cite{Happel2010}. As a consequence, a point-by-point tracing of the $E_{\rm r}$ measurement points from the SOL to the divertor or the calculation of the radial derivative of $E_{\rm r}$ is not fully reliable. Thus, in this work, it is assumed that DR measurement points in the SOL lie on the line of sight of the antenna. This is a safe assumption considering that the refractive index in the SOL region does not cause the reflectometer beam to deviate significantly from the line of sight of the antenna, given the low local values of the density typically found in the island. This assumption has nevertheless been verified by Travis calculations.\\ 

In order to discuss the evolution of the SOL electric field, it is useful to define first an average value of the $E_{\rm r}$ in the SOL ($\overline{E}_{\rm r, SOL}$) as a uniform value along the measurement region and the variation of the radial electric field on the sign reversal ($\Delta E_{\rm r}$) as a proxy for the $E_{\rm r}$ shear. For the calculation of $\overline{E}_{\rm r, SOL}$, we take the average for points with $E_{\rm r}>0$. While the observed values of $E_{\rm r, SOL}$ are rather constant for most analysed discharges (as in plot \ref{subfigure:180920047_profileRadialE}), for plasmas with higher densities, the cut-off position moves outwards measuring a substantial decline of $E_{\rm r}$ across the SOL (example in plot \ref{subfigure:180905017_profileRadialE}). In order to calculate $\overline{E}_{\rm r, SOL}$ in the most uniform manner, we use points close to the $E_{\rm r}$ sign reversal and ignore the ones corresponding to the referred decline. This is indicated in the examples in figure \ref{figure:betterProfileEr} by the colored markers in each plot: For 20180905.017, $\overline{E}_{\rm r, SOL}$ was found to be around $9 {\rm kV/m}$ and for 20180920.047, $\overline{E}_{\rm r, SOL} \simeq 10 {\rm kV/m}$; depicted in the two figures as a dashed black horizontal line. $\Delta E_{\rm r}$ is calculated as the difference between the values of the radial electric field on the sign reversal. It is found to be approximately $15 \rm kV/m$ for 20180905.017 and $14 \rm{kV/m}$ for 20180920.047 as indicated by a double vertical arrow in the two figures. \\

\begin{figure}
    \centering
    \setbox1 = \hbox{\includegraphics[height=15cm]{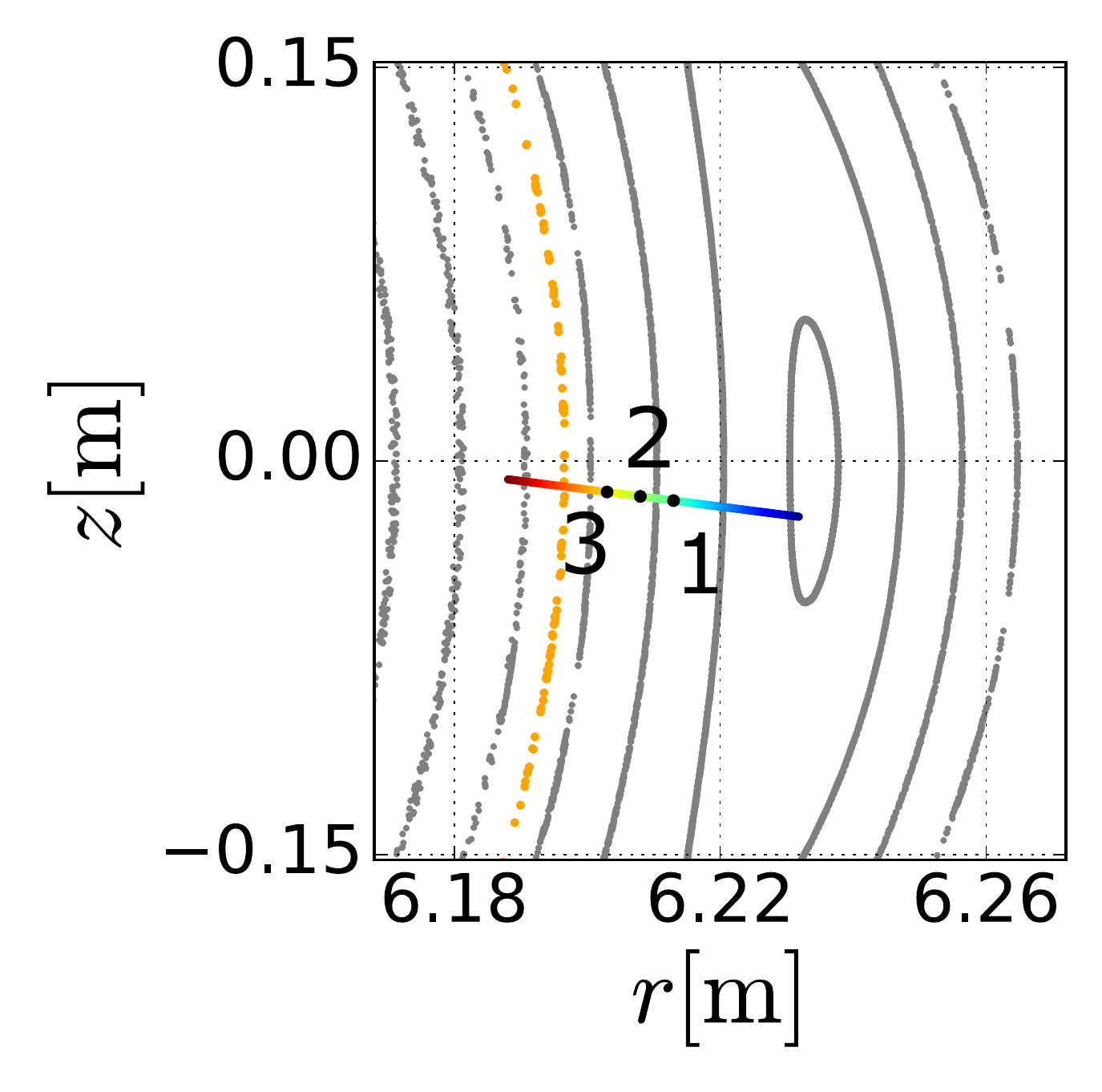}}
    \includegraphics[height=11cm]{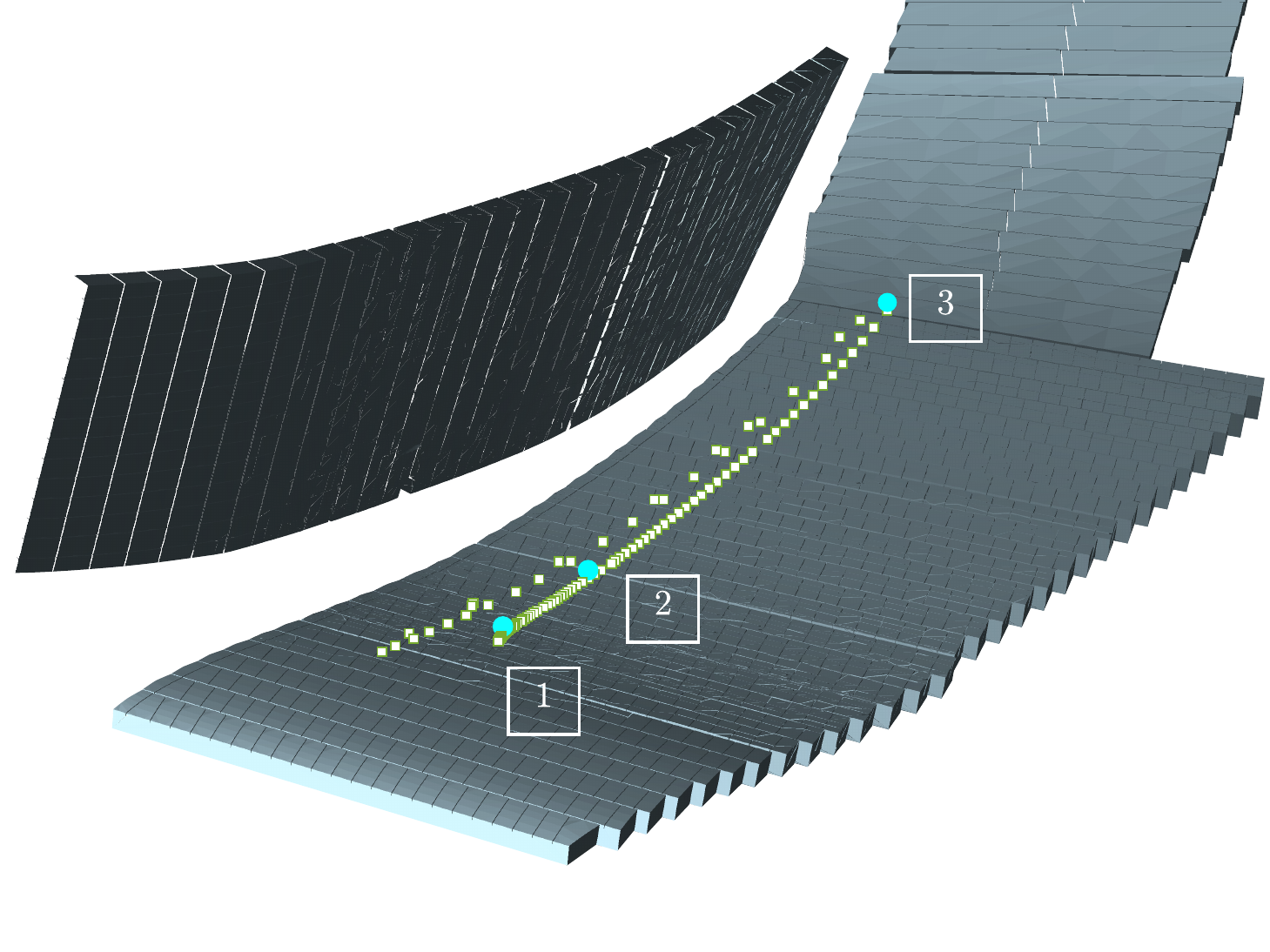}\llap{\makebox[\wd1][l]{\raisebox{7.5cm}{\includegraphics[height=5.0cm]{images/experimentalSetUp/180905017_4750_5000_poincare_zoom_markers_small.pdf}}}}
    \caption{The ending points of the field line tracing with starting points from the measurement area from the upstream (seen in the supplementary figure) to the divertor target plates. The ending points lie over the 2D surface of the s1lh target plate. For easier visualization of this projection from the SOL to the target numbered markers are used between the two plot. In the supplementary figure, the poloidal cross-section of $5/5$ island of the standard configuration for program 20180905.017 ($t = 4,750 \rm ms$ to $5,000 \rm ms$) with the measurement area of the DR as colored line.}
    \label{figure:complexProjection}
\end{figure}

An important part for this analysis is the connection of the measurement region in the SOL (plot \ref{subfigure:methodPoincareZoom}) with the divertor targets of the device along the magnetic field lines. The magnetic field lines are calculated using the aforementioned FLT. Aside the used currents of the NPC's and PC's, any non-negligible control coil current ($I_{\rm{CC}}$) and net toroidal current in the plasma ($I_{\rm tor}$) were also considered in the calculation of the magnetic field. In particular, $I_{\rm tor}$ was modelled as a current filament along the magnetic axis of the standard configuration. In order to make this calculation meaningful, discharges and times have been selected such that $I_{\rm tor}$ can be considered constant during one full DR frequency scan. For example, in the reference experiment 20180905.017, for $4,750{\rm ms}-5,000{\rm ms}$  $I_{\rm tor}$ has an average value of approximately $3.2 \rm kA$ with standard deviation that is $\sim 40 \rm{A}$. In plot \ref{figure:complexProjection}, the s1lh and s1lv divertor target plates are depicted. The green markers that lie on its surface represent the projection of the measurement area in the SOL. For easier identification and understanding of the projection of the measurement area on the target, three numbered markers are used for the two plots. As will be seen later, those markers also indicate the region of interest for the calculation of important parameters on the divertor. The ending points of the field line tracing lie all over the 2D surface of the target and do not align with the radial or toroidal direction in a simple way. Because of this, measurements of divertor parameters are required across the whole the surface of the divertor plate in order to ensure an overlap between the projection from the upstream and the measurement area of diagnostics for the divertor. This complicates substantially the analysis, as it renders that probe systems on the target impractical for this analysis, as they are arranged in an array extending across the divertor along the poloidal direction \cite{Laube2011} (figure 1 therein). \\


\begin{figure}[!ht]
  \centering
  \includegraphics[width=0.75\textwidth]{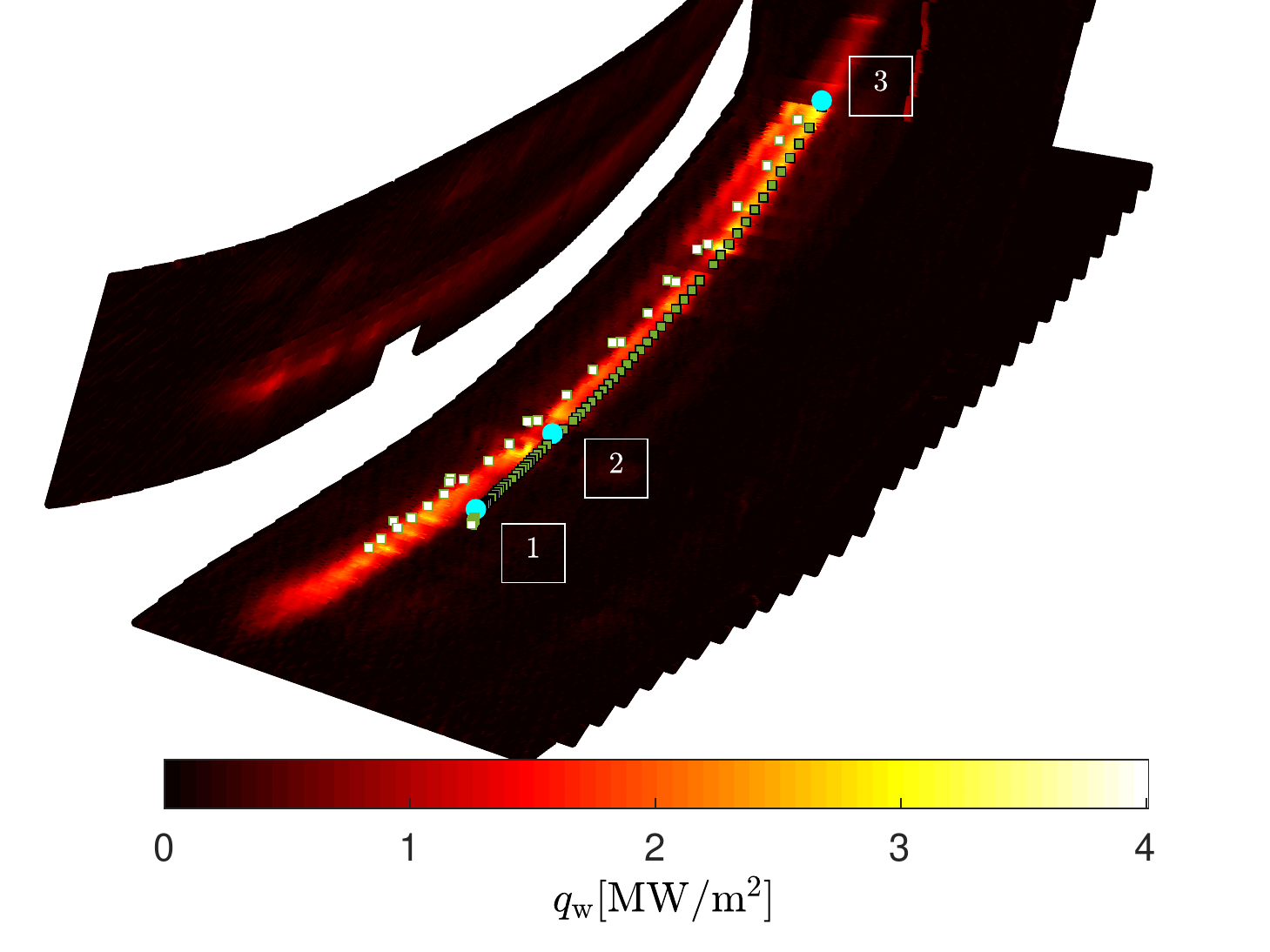}
\caption{For program 20180905.017 ($t = 4,750 \rm ms$ to $5,000 \rm ms$) the heat fluxes $q_{\rm w}$ onto the target s1lh along with the projection of the measurement area with square markers.}
\label{figure:heatIR}
\end{figure}

\begin{figure}
    \centering
    \includegraphics[width=0.49\textwidth]{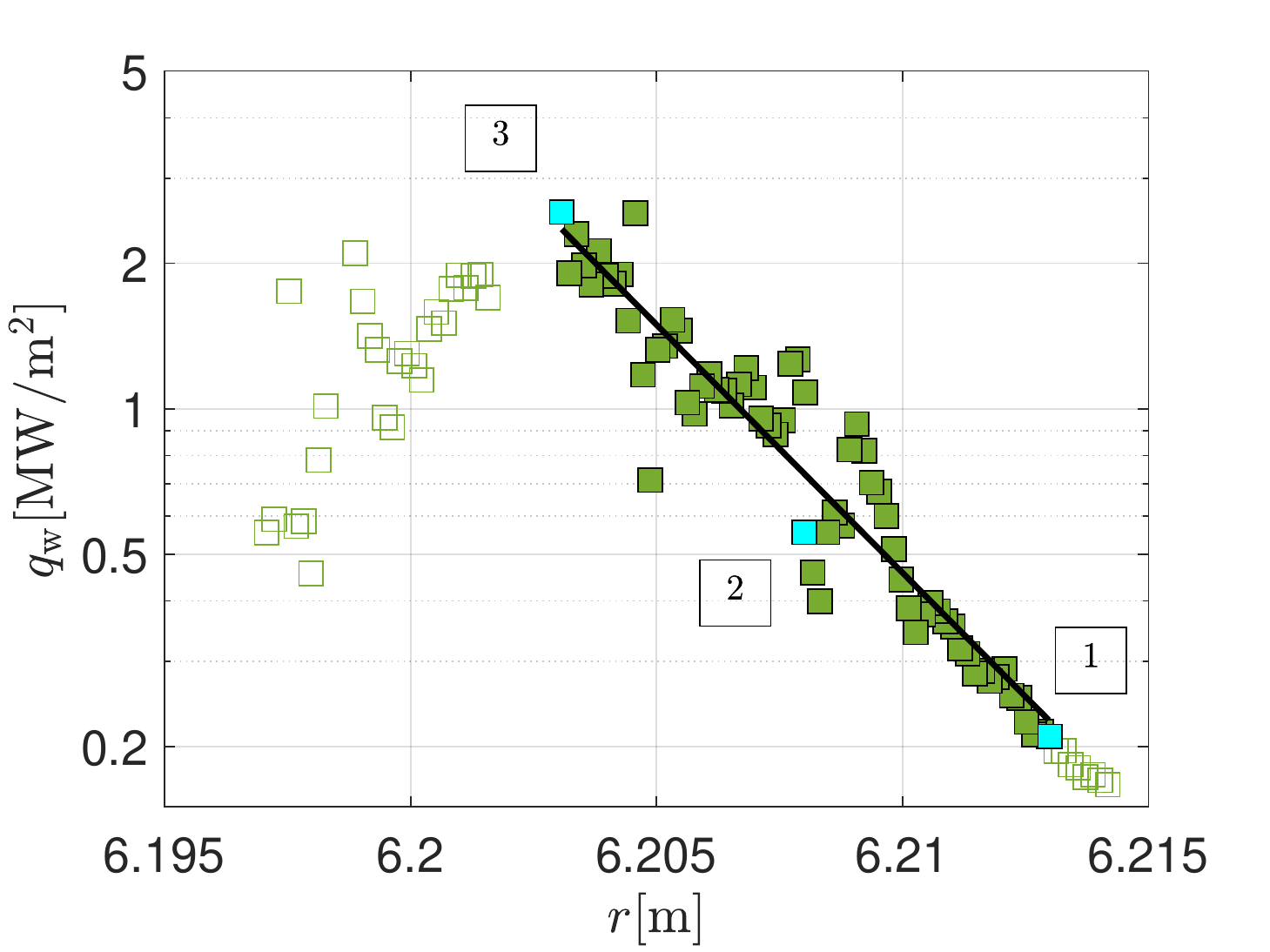}
    \caption{Heat fluxes for each point on the divertor from plot \ref{figure:heatIR} for W7-X program 20180905.107 ($t = 4,750 \rm ms$ to $5,000 \rm ms$). Hollow markers correspond to points that were not used for the fit and the calculation of $\lambda_{q}$ through the linear fit (solid black line). The three cyan numbered markers correspond to the numbered markers of figure \ref{figure:complexProjection}.}
    \label{figure:heatProfile}
\end{figure}

Given the need for a full 2D diagnostic, we resort to the infrared camera systems installed in W7-X \cite{Gao2019}: This diagnostic captures thermographic images of the full divertor surface thus covering the projected area from the upstream measurements. Those images are used as an input for THEODOR code \cite{Herrmann1995}, which calculates the heat fluxes that reach the target wall ($q_{\rm w}$). Typically, IR data cannot be reliably used for this analysis if the ratio of $P_{\rm rad}/P_{\rm ECH} > 0.5$, where $P_{\rm rad}$ is the power radiated by the plasma, due to the low signal to noise ratio. This is limiting our study mainly to not detached plasmas. In plot \ref{figure:heatIR}, we show the heat flux onto target s1lh for W7-X program 20180905.017 ($t = 4,750 \rm ms$ to $5,000 \rm ms$) as an example. In plot \ref{figure:heatIR}, is also seen the projection of the measurement area is included with square markers. We see that our projection from the SOL lies close to the strike point, around the region where the most intense heat fluxes are observed. It is now possible to extract the heat flux onto the divertor for its corresponding measurement point in the SOL, which is displayed in plot \ref{figure:heatProfile}. If $T_i \simeq T_e$ is assumed in the vicinity of the divertor, ion reflection is considered negligible (which is typically the case for a carbon target) and the usual sheath entrance condition $v_s = c_s$ is taken, the heat flux entering the sheath in front of the wall, $q_{\rm t}$ can be expressed in terms of the plasma parameters as in \cite{stagenbybook}:

\begin{equation}
    q_{\rm t} = c_s n_{c,s}[\gamma_s T_{e,s} + E_{\rm rec}]
    \label{eq:heatFluxInitial}
\end{equation}
where $c_s$ is the isothermal sound speed at the entrance of the sheath, $c_s = \sqrt{2T_e / m_i}$ and $\gamma_s \simeq 7.5$ is a sheath transmission coefficient. $E_{\rm rec}$ is the energy delivered to the target due to the atomic processes (including both ionization and dissociation energies), which can be considered constant and will therefore not matter for the discussion of this work \cite{Brida2017}. Using the process previously described, it is possible to use the heat fluxes obtained from IR measurements to evaluate the decay of the heat flux over the DR measurement projection and use it to provide an estimation of the temperature decay in order to evaluate the relation described in equation \ref{eq:finalEr}. In particular, if an exponential decay is assumed, $q_{\rm t} = q_{0}\rm{exp}(-r / \lambda_{\rm q})$, the exponential decay length $\lambda_{\rm q}$ can be computed. Indeed, if a similar exponential decay is assumed for the density, for weak parallel gradients in the SOL, the radial derivative of equation \ref{eq:heatFluxInitial} at the mid-plane becomes $\lambda_{\rm q}^{-1} = \lambda_n^{-1} + \frac{3}{2}\lambda_T^{-1}$. By taking $\lambda_n \simeq \lambda_T$ (which seems reasonable looking eg. at the target profiles from \cite{Hammond2019}), this can be further simplified to $\lambda_{\rm q} \simeq \frac{2}{5} \lambda_T$. In the conduction limited regime, featuring strong parallel gradients, the Two-Point model relation $\lambda_{\rm q} \simeq \frac{2}{7} \lambda_T$ can be used instead \cite{stagenbybook}. This means that the two e-folding lengths are probably such that $\frac{2}{7} < \lambda_{\rm q} / \lambda_{T_{e}} < \frac{2}{5}$, and $\lambda_{\rm q} \propto \lambda_T$ in any of the two cases. It must be taken into account that these $q_{\rm t}$ and $\lambda_{\rm q}$ refer heat flux entering the sheath, while the IR system measures the heat flux perpendicular to the target wall, $q_{\rm w}$. The two quantities differ due to the grazing angle ($\alpha$) between the magnetic field lines and the divertor target plates $q_{\rm w} = q_{t} sin(\alpha)$ \cite{Gao2020}. However, $\alpha$ is not found to change significantly at the relevant projection region of the divertor for the standard configuration, so it can be considered that $q_{\rm w} \propto q_{t}$ all across the region where $\lambda_{\rm q}$ is calculated. Thus, $\lambda_{\rm q}$ can be used as a qualitative proxy for the temperature decay length in equation \ref{eq:finalEr}. \\

In order to carry out the analysis, certain criteria must be established for the points that will be used for the calculation of the the referred exponential decay length. The starting point for the fit should be considered the point with the maximum heat flux (see marker 3 on figure \ref{figure:heatProfile}), including points for increasing values of $r$. The second criterion is related to the minimum meaningful value of $q_{\rm w}$, below which points should not be considered. Points with $q_{\rm w} < 0.20 {\rm MW/m^{2}}$ are too close to the background noise level and excluded since the quality of the IR measurements is considered insufficient for them (see marker 1 and for increasing $r$ the hollow markers on figure \ref{figure:heatProfile}). As well, points that are too close to the LCFS should be excluded. In particular, it has been determined that points with $r$ less than $\sim 6.2 {\rm m}$ should be excluded from the fit since the total connection length ($L$) increases very rapidly in a radial region of an extension below the spatial resolution of the field line tracing code. A representative example of this (corresponding to discharge 20180905.017) can be seen in figure  \ref{subfigure:connectionLength}, where $L$ values diverge over $350\rm{m}$ for $r < 6.2 {\rm m}$. Since we cannot know whether those points are indeed outside or inside the LCFS, they are excluded from the linear fit and depicted as hollow markers on the left of figure \ref{figure:heatProfile}. Finally, those remaining points that are eligible for the fit are shown as markers filled with green color. The electron temperature for the ratio on the rhs of \ref{eq:finalEr} is calculated through the TS temperature profiles at the last closed flux surface since low parallel gradients are assumed from the lcfs to the sheath entrance. \\
 
\begin{figure}[!ht]
\centering
\begin{subfigure}{0.49\textwidth}
  \includegraphics[width=\linewidth]{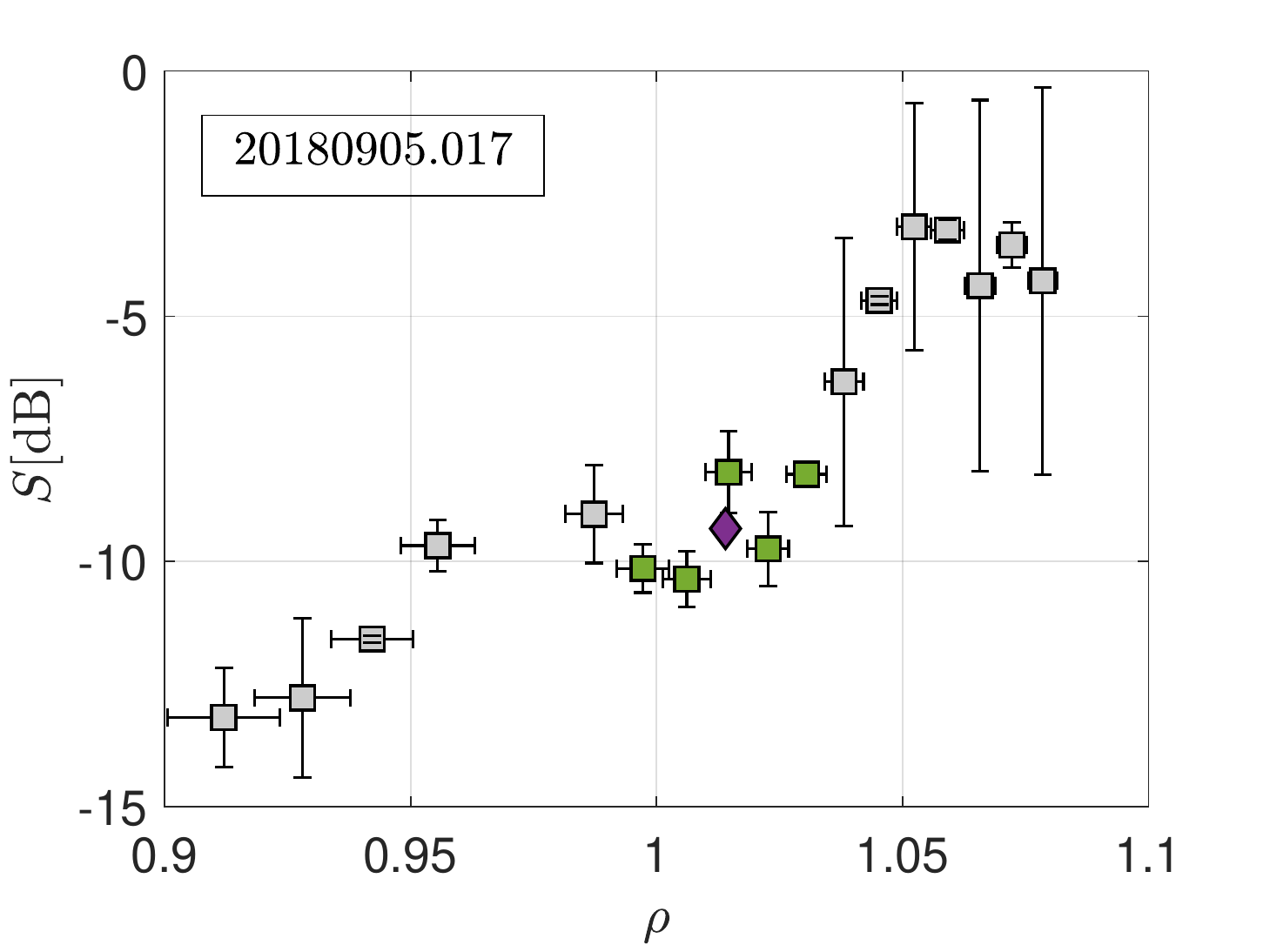}
  \caption{}
  \label{subfigure:20180905017_profileFluct}
\end{subfigure} \hfil
\begin{subfigure}{0.49\textwidth}
  \includegraphics[width=\linewidth]{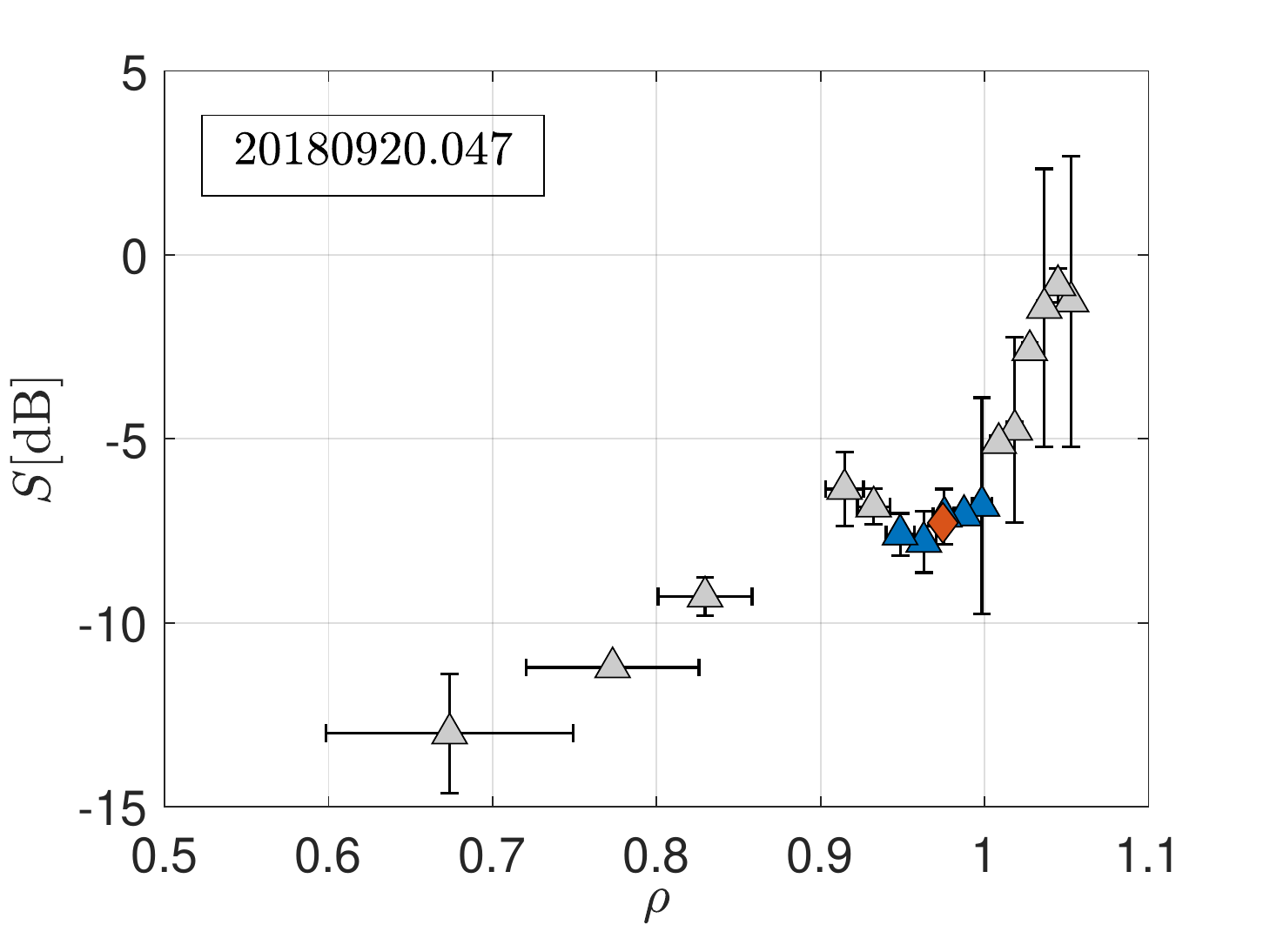}
  \caption{}
  \label{subfigure:20180920047_profileFluct}
\end{subfigure}
\caption{$S$ profiles by the DR for discharge $20180905.017$ ($t = 4,750\rm{ms}$ to $t=5,000\rm{ms}$) and for discharge $20180920.047$ ($t = 7,000\rm{ms}$ to $t=7,250\rm{ms}$).}
\label{figure:exampleSuppression}
\end{figure}

As mentioned in the Introduction, the other main objective of this work is to seek any trends between $\Delta E_{\rm r}$ (as defined in figure \ref{figure:betterProfileEr}) and the amplitude of edge turbulence. For this task, we used the density fluctuation measurements from the DR. In plots \ref{subfigure:20180905017_profileFluct} and \ref{subfigure:20180920047_profileFluct}, the $S \propto \delta n_{\rm e}^2$ profiles are shown for values around $\rho = 1$. For both discharges, approximately at the same position as the $E_{\rm r}$ sign reversal, a local minimum of the power collected by the DR is observed for a number of measurement points, as indicated by the colored markers in figures \ref{subfigure:20180905017_profileFluct} and \ref{subfigure:20180920047_profileFluct}. By taking the average value of $S$ and $\rho$ of points that constitute this local minimum, it is possible to define the average reduced level of power collected by the DR ($\overline{S}_{\rm min}$, as the diamond marker with distinct color on each figure). For example, in the previously discussed cases $20180905.017$ and $20180920.047$, $\overline{S}_{\rm min}$ is found approximately $-9.3\rm{dB}$ and $-7.3\rm{dB}$ respectively, with an uncertainty of the order of $2 \rm{dB}$. This approach will be used in section \ref{section:results} for a bigger list of experiments, in or between $\Delta E_{\rm r}$ and $\overline{S}_{\rm min}$. \\

\label{section:experimental_setup}

\newpage 

\section{Results}

\begin{figure}[!ht]
\centering
\includegraphics[scale=0.49]{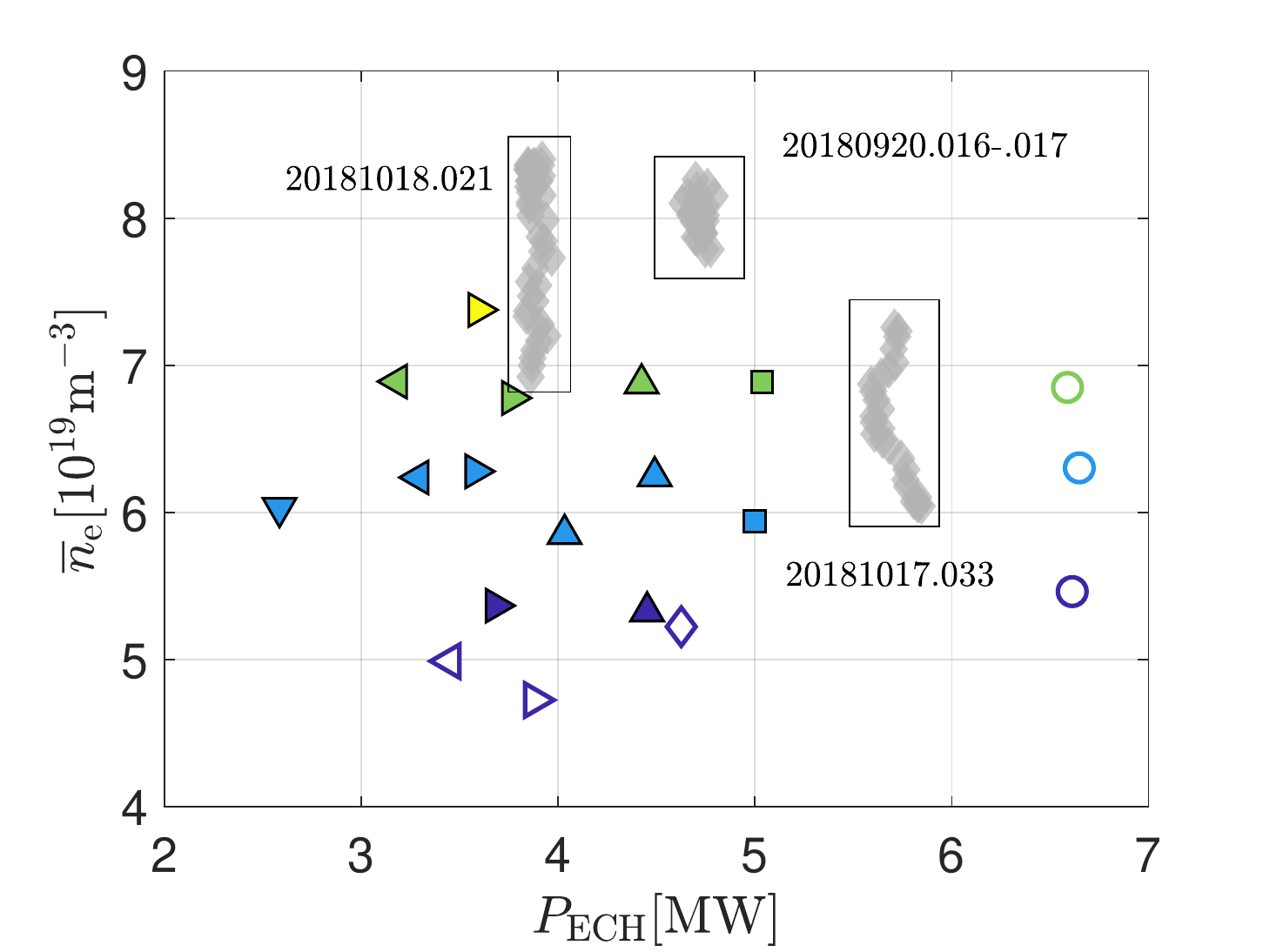}
\caption{Parametric map with all analysed discharges displayed as a function of the ECH power ($P_{\rm ECH}$) and line averaged density ($\overline{n}_{\rm e}$) that are considered in this work. Grey markers and hollow markers depict discharges that will be used for the study of the edge $E_{\rm r}$ shear and turbulence reduction but not for the study on the formation of the $E_{\rm r, SOL}$.}
\label{figure:densityECHMaps}
\end{figure}

For the purposes of this study, a set of various plasmas heated with electron cyclotron heating (ECH) was analysed. The standard configuration was preferred over other configurations due to the fact that it was used most commonly during OP1.2. In the selected discharges, that are presented in a parametric map on figure \ref{figure:densityECHMaps}, we have explored the available range of density and heating powers that could guarantee the existence of measurement points of the DR in the SOL. For each point of this figure a full DR frequency scan was analysed for the indicated times. Colors on this plot indicate different levels of the line averaged electron plasma density, $\overline{n}_{\rm e}$. Instead, different markers indicate similar values of ECH power. It can be seen that the range of line averaged densities varies from $4.5 \times 10^{19} \rm m^{-3}$ to $7.5 \times 10^{19} \rm m^{-3}$ which covers the plasma scenarios for which V-band frequencies with O-mode polarization can measure in the SOL region. The level of ECH power varies from $2.5 \rm MW$ and $6.5 \rm MW$, the latter value being close to the maximum heating power coupled to the plasma in W7-X \cite{Erckmann2017, Wolf2017}. For the formation of $E_{\rm r, SOL}$, we studied plasmas with density from $\simeq 5.5 \times 10^{19} \rm m^{-3}$ to $7.5 \times 10^{19} \rm m^{-3}$ and ECH power from $2.5 \rm MW$ to $5 \rm MW$. In table \ref{table:dischargesDivertor}, we list the analysed discharges along with the DR scan time intervals, the control coil currents and net toroidal plasma currents that we used as input for the field line tracing calculation. Hollow markers represent discharges for which the heat fluxes on the target are not available. Nevertheless, they are still included in our database since they will be used for the study of $E_{\rm r, SOL}$ by other plasma parameters (i.e. $P_{\rm ECH}$ or $\overline{n}_{\rm e}$) and the study for $\Delta E_{\rm r}$ and the reduced $S$. Discharges with grey markers concern also the $\Delta E_{\rm r}$ and the reduced $S$ that will be discussed later. \\

\begin{table}[!ht]
\centering
\begin{tabular}{||c c c c c||} 
\hline
Discharge number & DR scan times $\rm{[ms]}$ & $I_{\rm CC} \rm [\rm kA]$ & $I_{\rm tor} \rm{[kA]}$ & Marker\\ [0.5ex] 
\hline\hline
20180807.022 & 4,000-4,252 & 0 & 1.84    & \myemptylefttriangle{mydarkblue} \\ 
20180814.042 & 1,000-1,252 & 0 & $\ll 1$ & \myemptyrighttriangle{mydarkblue} \\ 
20181010.012 & 2,750-3,001 & 0 & 1.93    & \myemptydiamond{mydarkblue} \\ 
20180807.022 & 1,250-1,502 & 0 & $\ll 1$ & \myfilledrighttriangle{mydarkblue} \\
20181010.012 & 1,000-1,252 & 0 & $\ll 1$ & \myfilleduptriangle{mydarkblue} \\ 
20180906.008 & 2,000-2,251 & 1 & 2.53    & \myemptycircle{mydarkblue} \\ 
20180807.017 & 2,000-2,252 & 0 & $\ll 1$ & \myfilleddowntriangle{mylightblue} \\ 
20180814.010 & 1,000-1,252 & 2 & $\ll 1$ & \myfilledlefttriangle{mylightblue} \\ 
20180807.020 & 5,500-5,752 & 0 & $\ll 1$ & \myfilledrighttriangle{mylightblue} \\
20180920.047 & 7,000-7,252 & 0 & 2.51    & \myfilleduptriangle{mylightblue} \\
20181010.007 & 8,500-8,752 & 0 & 3.89    & \myfilleduptriangle{mylightblue} \\
20180905.016 & 1,250-1,502 & 2 & 1.25    & \myfilledsquare{mylightblue} \\
20180906.008 & 4,250-4,502 & 1 & 4.38    & \myemptycircle{mylightblue} \\
20180905.015 & 10,500-10,753 & 0 & 1.71  & \myfilledlefttriangle{mygreen} \\
20180814.011 & 4,500-4,751 & 0 & $\ll 1$ & \myfilledrighttriangle{mygreen} \\
20181010.019 & 5,750-6,001 & 0 & 2.70    & \myfilleduptriangle{mygreen} \\
20180905.017 & 4,750-5,002 & 1 & 3.21    & \myfilledsquare{mygreen} \\
20180906.008 & 6,250-6,501 & 1 & 5.58    & \myemptycircle{mygreen} \\
20180807.020 & 1,250-1,502 & 0 & $\ll 1$ & \myfilledrighttriangle{myyellow} \\ [1ex]
\hline
\end{tabular}
\caption{Showing discharge numbers, the DR scan time intervals, $I_{\rm CC}$ and $I_{\rm tor}$ of the ECH plasmas used for the given analysis. Additionally, the equivalent marker of each DR scan for figure \ref{figure:densityECHMaps} is given in the last column. All discharges employ the standard configuration of W7-X. When $I_{\rm tor} \ll 1 \rm{kA}$, its value was considered negligible.}
\label{table:dischargesDivertor}
\end{table}

\subsection{Impact of the divertor on SOL radial electric field}

\begin{figure}[!ht]
    
    \centering
    \subfloat[]{
    \label{subfigure:radialEdensity}
    \includegraphics[width=0.49\textwidth]{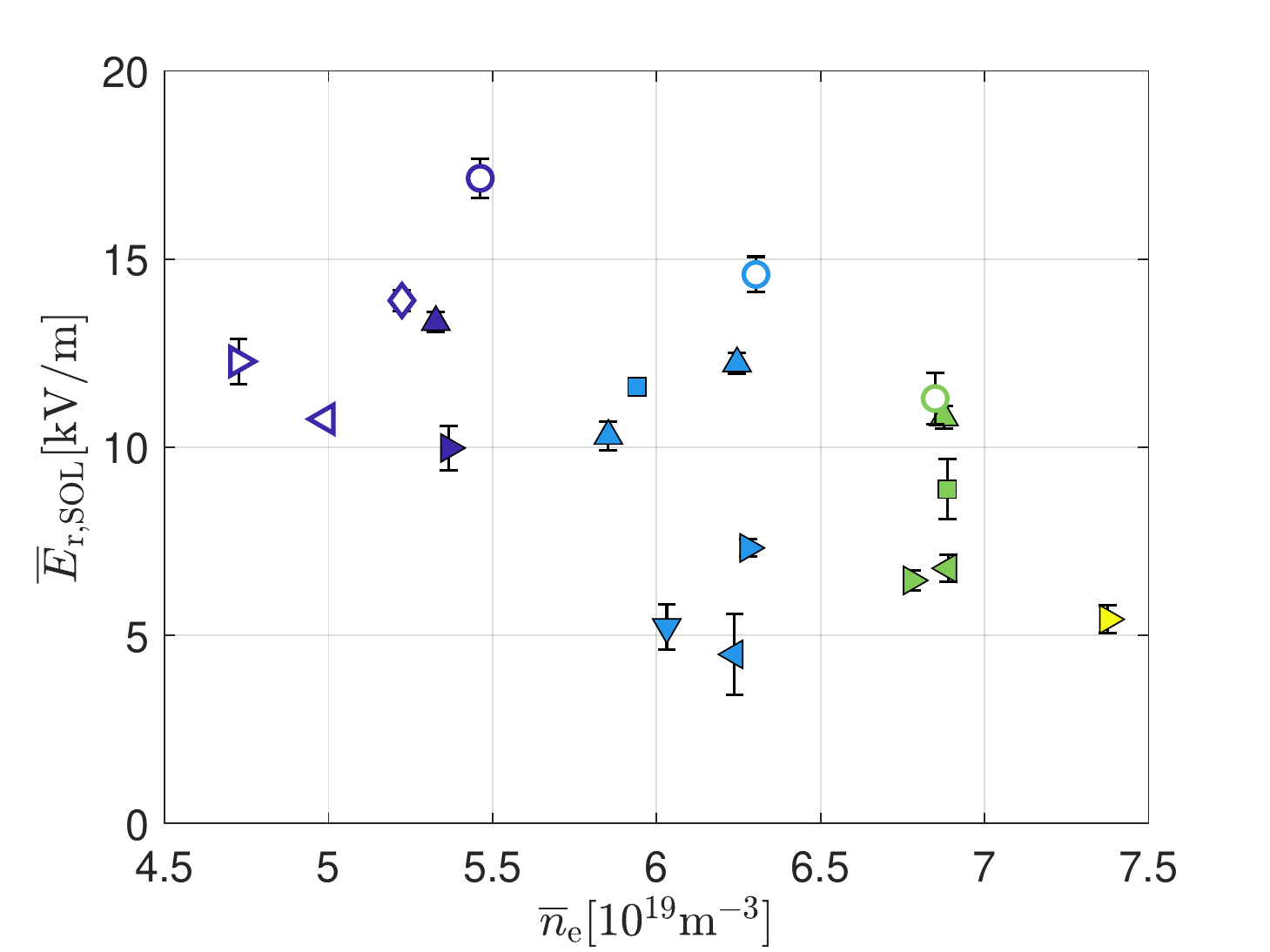}}
    
    \subfloat[]{
    \label{subfigure:radialEECH}
    \includegraphics[width=0.49\textwidth]{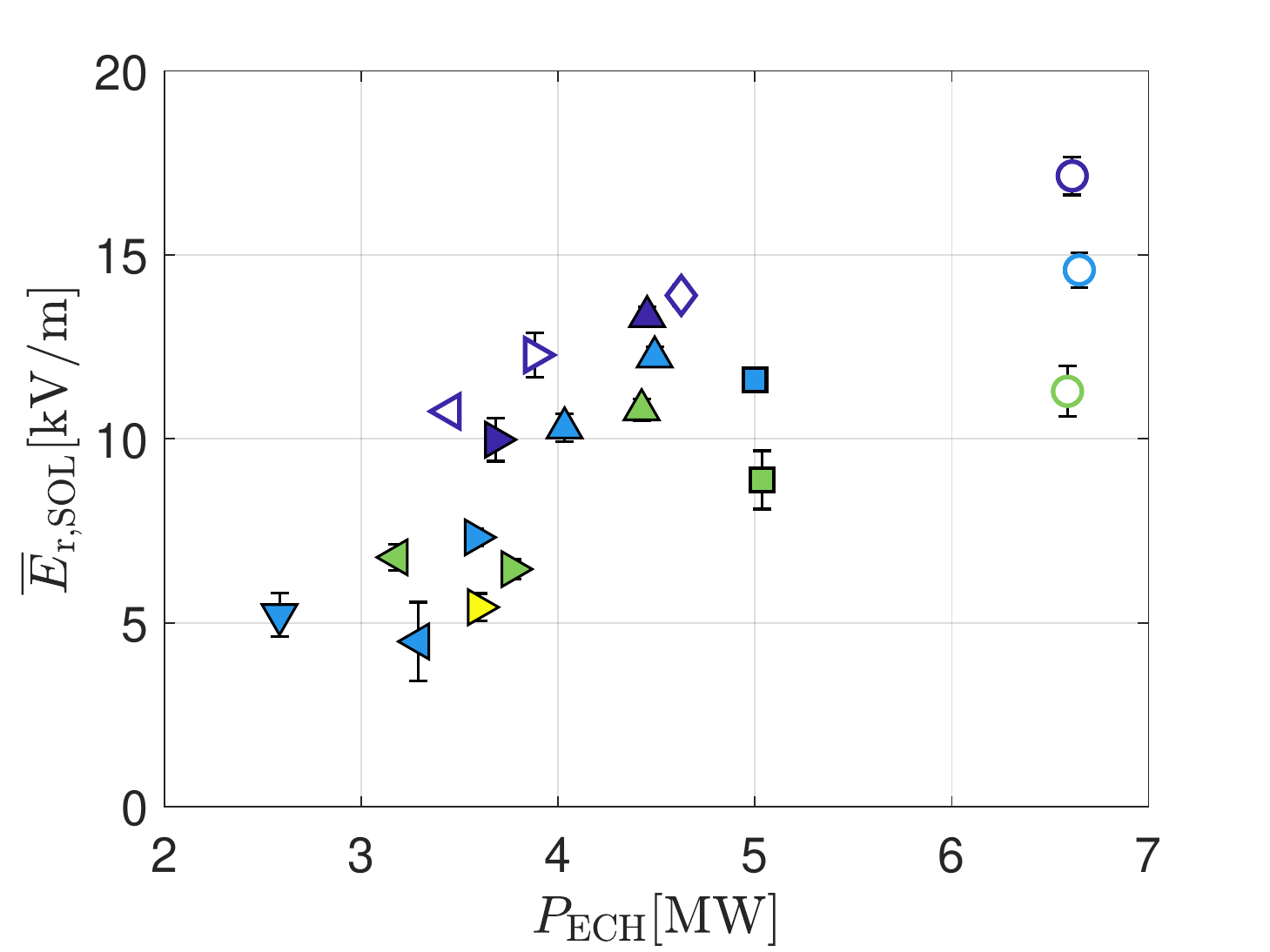}}
    
    \subfloat[]{
    \label{subfigure:echDensityRadialE}
    \includegraphics[width=0.49\textwidth]{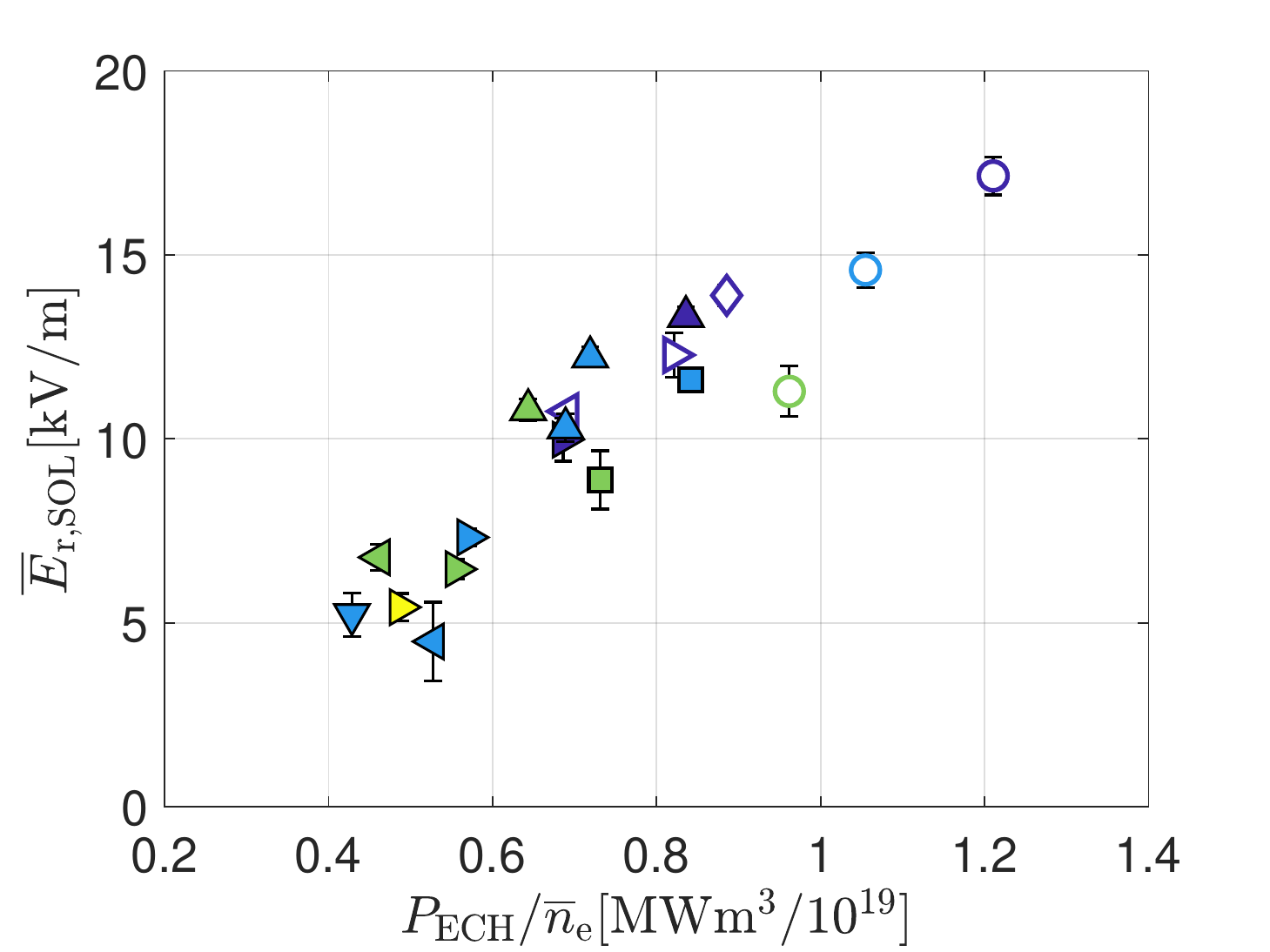}}
    
    \caption{$\overline{E}_{\rm r,SOL}$ as a function of $\overline{n}_{\rm e}$ (figure \ref{subfigure:radialEdensity}), $P_{\rm ECH}$ (figure \ref{subfigure:radialEECH}) and the $P_{\rm ECH} / \overline{n}_{\rm e}$ ratio (figure \ref{subfigure:echDensityRadialE}). The use of markers and colors is the same as in figure \ref{figure:densityECHMaps}}
    \label{figure:correlationsAlpha}

\end{figure}

First, we study the evolution of $\overline{E}_{\rm r, SOL}$ (calculated as explained in section \ref{section:experimental_setup}) with respect to basic plasma parameters, like $\overline{n}_{\rm e}$ and $P_{\rm ECH}$. In plots \ref{subfigure:radialEdensity} and \ref{subfigure:radialEECH}, $\overline{E}_{\rm r, SOL}$ is plotted against the $\overline{n}_{\rm e}$ and $P_{\rm ECH}$ respectively. In plot \ref{subfigure:radialEdensity}, for increasing line averaged density and constant ECH power (see markers of similar shape but different color), the value of the radial electric field in the SOL region is decreasing. In plot \ref{subfigure:radialEECH}, for increasing ECH power and constant density (see markers of similar color but different shape), the radial electric field in the SOL is increasing. This result is consistent with previous reports \cite{Carralero2020}, covering a shorter range of ECH power values. From plots \ref{subfigure:radialEdensity} and \ref{subfigure:radialEECH}, it could be assumed that there is a proportional relationship of $\overline{E}_{\rm r, SOL}$ to ECH power and inversely proportional to the line density. This is confirmed in plot \ref{subfigure:echDensityRadialE} where $\overline{E}_{r, SOL}$ is plotted against the ratio of $P_{\rm ECH}/ \overline{n}_{\rm e}$. As can be seen, a rather linear relation exists between both parameters for the whole set of data, suggesting $\overline{E}_{r, SOL} \propto P_{\rm ECH}/ \overline{n}_{\rm e}$. The scattering of the points in plot \ref{subfigure:echDensityRadialE} is reduced with respect to the scattering of the points in plots \ref{subfigure:radialEdensity} and \ref{subfigure:radialEECH}. Thus, this ratio seems to set a much clearer linear trend in the value of the mid-plane electric field than solely $\overline{n}_{\rm e}$ or $P_{\rm ECH}$. \\

\begin{figure}[!ht]
\centering
\includegraphics[width=0.49\textwidth]{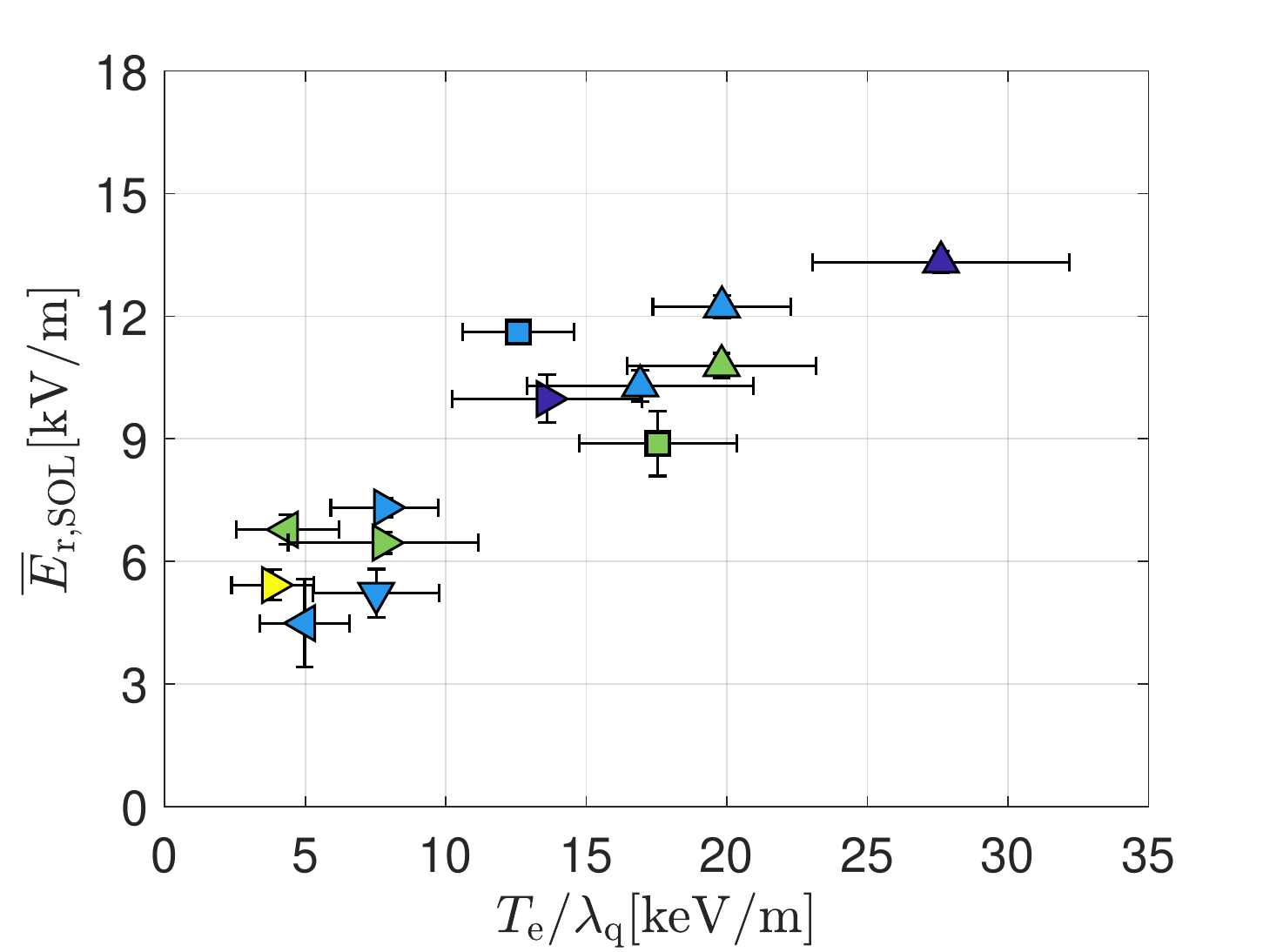}
\caption{$\overline{E}_{\rm r, SOL}$ and the ratio of the electron temperature with the exponential decay length of heat fluxes at the target, $T_{\rm e} / \lambda_{\rm q}$. The use of markers and colors is the same as in figure \ref{figure:densityECHMaps}.}
\label{figure:lambdaRadialE}
\end{figure}

Next, we evaluate the influence of the downstream divertor conditions on the mid-plane radial electric field. In figure \ref{figure:lambdaRadialE}, we plot $\overline{E}_{\rm r, SOL}$ and the ratio of the electron temperature over the exponential decay length of the heat flux, $T_{\rm e} / \lambda_{\rm q}$. The plotted data shows that $\overline{E}_{\rm r, SOL}$ scales clearly with the ratio $T_{\rm e} / \lambda_{\rm q}$, roughly verifying for the whole database the relation described in equation \ref{eq:finalEr}. Despite the approximations discussed in the previous sections, the radial electric field in the SOL appears to be dominated by the conditions on the divertor for the studied plasmas. The uncertainty bars of the radial electric field are computed as the standard deviation of the points that we used to compute the mean average value. The uncertainty bars of $\lambda_{q}$ were calculated from linear regression on the points of figure \hyperref[figure:heat]{3c}. For the majority of discharges, the relevant divertor target was s1lh. However, for discharges 20181010.007 and 20181010.019 (both of them featuring high toroidal plasma currents, $3.9 \rm kA$ and $2.7 \rm kA$ respectively with no current in the control coils), the FLT analysis showed that the region relevant for the calculation of $\lambda_{\rm q}$ was found at the s4uv target. A mapping of different positions in the island onto the different divertor targets of W7-X has been shown in \cite{Killer2019}. Nevertheless, both groups of points seem to merge seamlessly in figure \ref{figure:lambdaRadialE}, indicating that the changes on the SOL magnetic field structure induced by the toroidal current do not significantly affect the general trend. \\

\subsection{Effect of $E_{\rm r}$ shear on the edge fluctuation level}

Since only DR data is required for the study of the impact of $\Delta E_{\rm r}$ on edge turbulence, it was possible to expand the existing database including other discharges with numerous scans of the reflectometer (96 additional DR scans from 4 more discharges). In table \ref{table:dischargesSuppression} we present the discharge numbers with the corresponding time intervals of the extra DR scans. As well, in figure \ref{figure:densityECHMaps}, these discharges are represented as grey points, to set them apart from the ones on which IR data is discussed. Interestingly, two of them (20181017.033 and 20181018.021) feature density ramps, on which $\overline{n}_{\rm e}$ is steadily increased while keeping constant the ECH power. Further details on these data can be found elsewhere in the literature \cite{Carralero2021}. \\

\begin{table}[!ht]
\centering
\begin{tabular}{||c c c c||} 
\hline
Discharge number & DR scan times $\rm{[ms]}$ & $\overline{n}_{\rm e} \rm [\rm 10^{19}m^{-3}]$ & $P_{\rm ECH}\rm{[MW]}$\\ [0.5ex] 
\hline\hline
20180920.016 & 2,250-6,000 & 6.2 & 4.5 \\ 
20180920.017 & 2,000-6,000 & 6.2 & 4.5 \\ 
20181017.033 & 2,750-9,500 & 4.6 - 5.8 & $< 6$ \\ 
20181018.021 & 1,000-12,000 & 5.4 - 6.5 & 4 \\ [1ex]
\hline
\end{tabular}
\caption{Table with discharge numbers, the given DR scans, $\overline{n}_{\rm e}$ and $P_{\rm ECH}$ of the ECH plasmas used for the given analysis. All discharges employ the standard configuration of W7-X.}
\label{table:dischargesSuppression}
\end{table}

\begin{figure}[!ht]
    
    \centering
    \subfloat[]{
    \label{subfigure:Swithdensity}
    \includegraphics[width=0.49\textwidth]{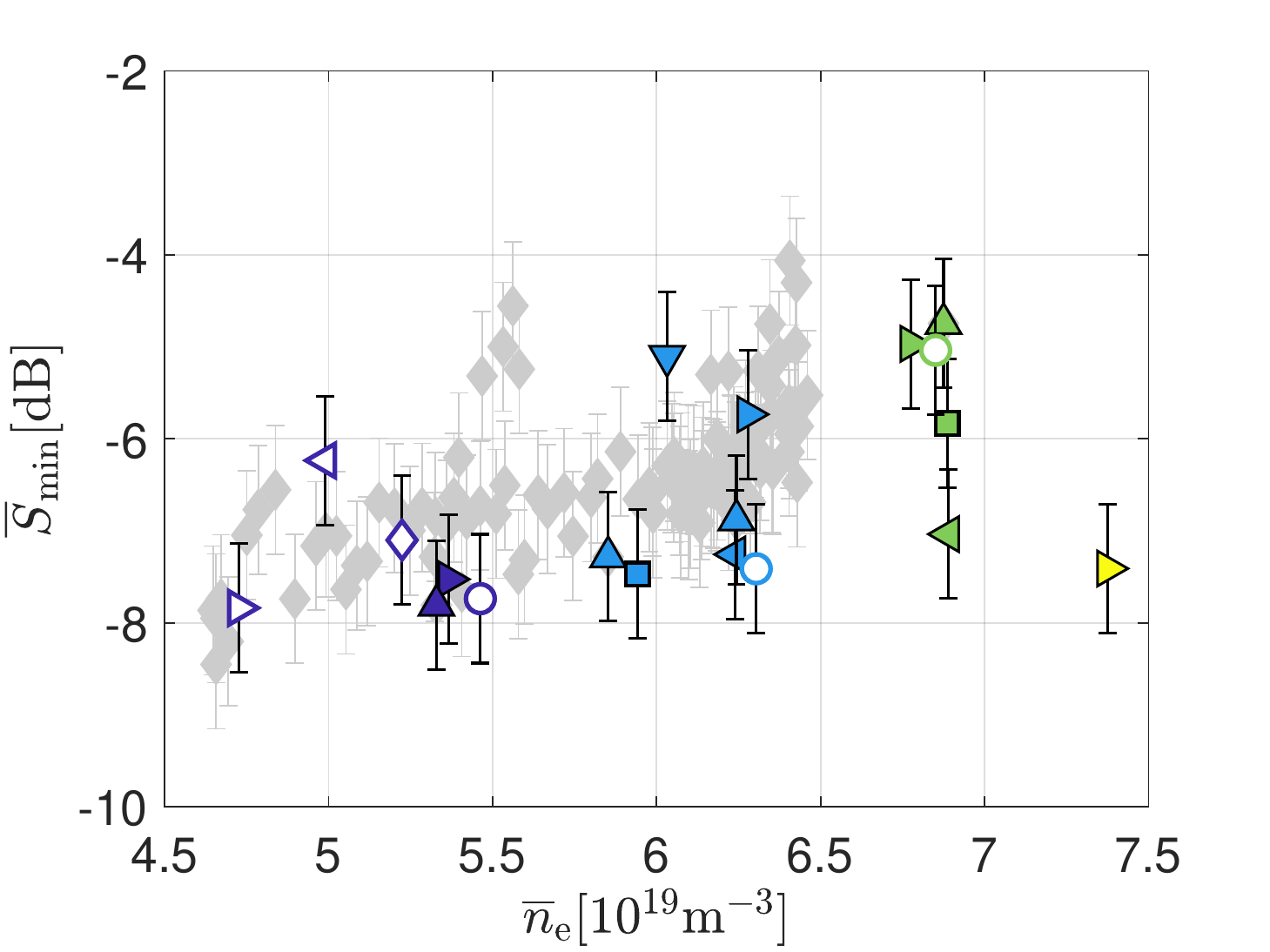}}
    
    \subfloat[]{
    \label{subfigure:SwithECH}
    \includegraphics[width=0.49\textwidth]{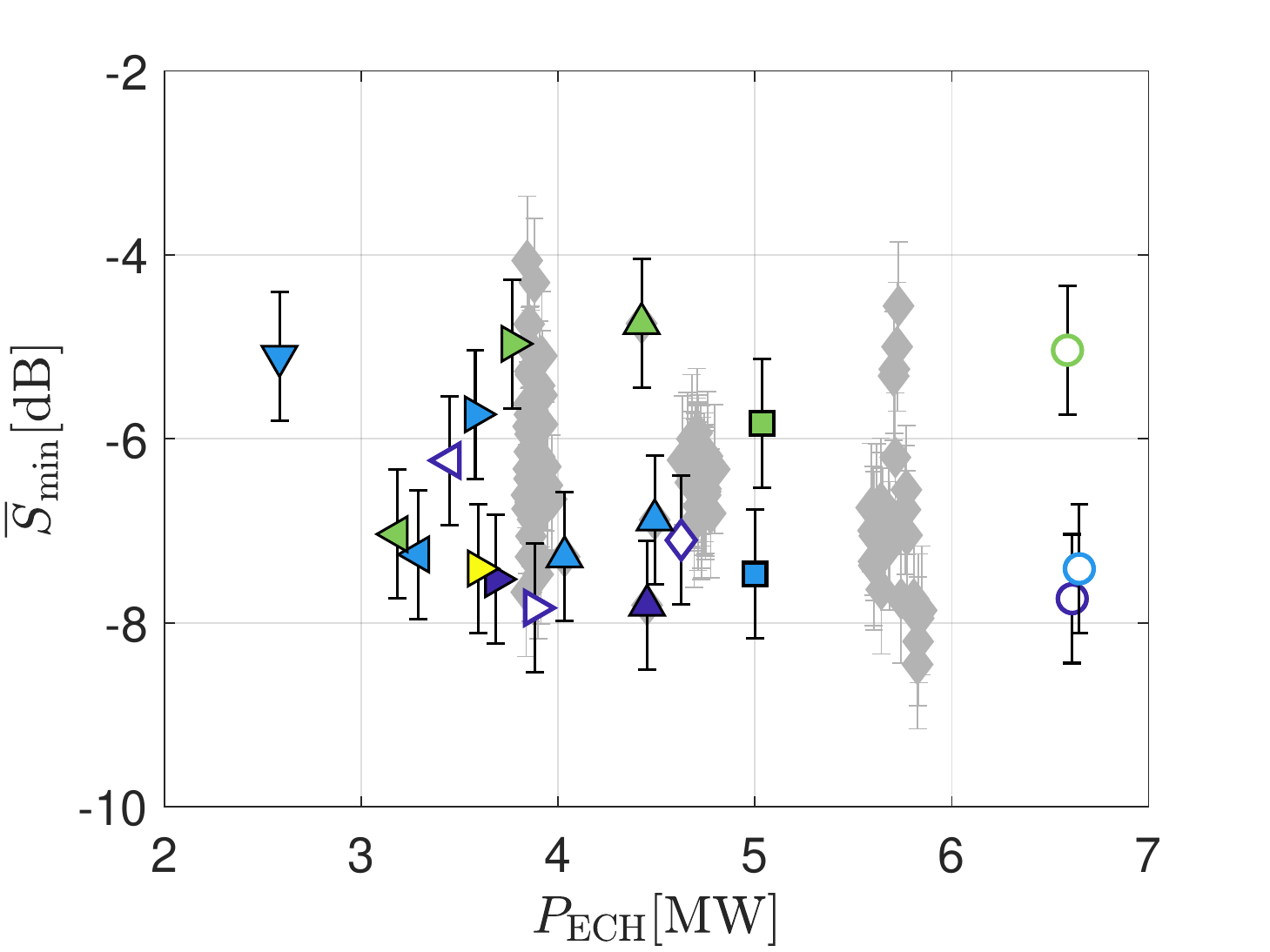}}
    
    \subfloat[]{
    \label{subfigure:SwithECHDensity}
    \includegraphics[width=0.49\textwidth]{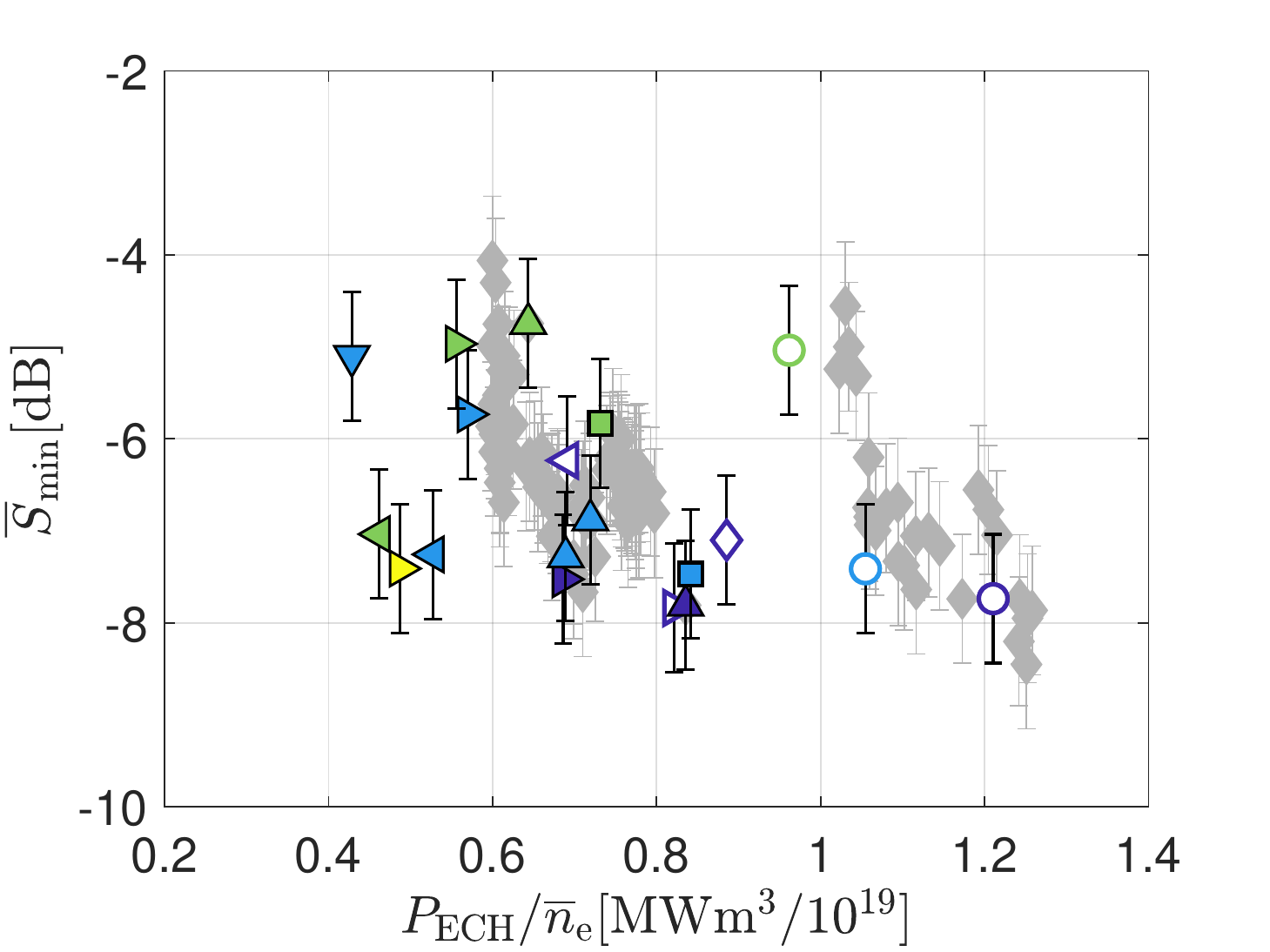}}
    
    \caption{The back-scattered power $S\propto \delta n_{\rm e}^2$ as a function of $\overline{n}_{\rm e}$ (figure \ref{subfigure:Swithdensity}), $P_{\rm ECH}$ (figure \ref{subfigure:SwithECH}) and the $P_{\rm ECH} / \overline{n}_{\rm e}$ ratio (figure \ref{subfigure:SwithECHDensity}). The use of markers and colors is the same as that was given in figure \ref{figure:densityECHMaps}.}
    \label{figure:SwithParameters}

\end{figure}

\begin{figure}[!ht]
\centering
\includegraphics[scale=.49]{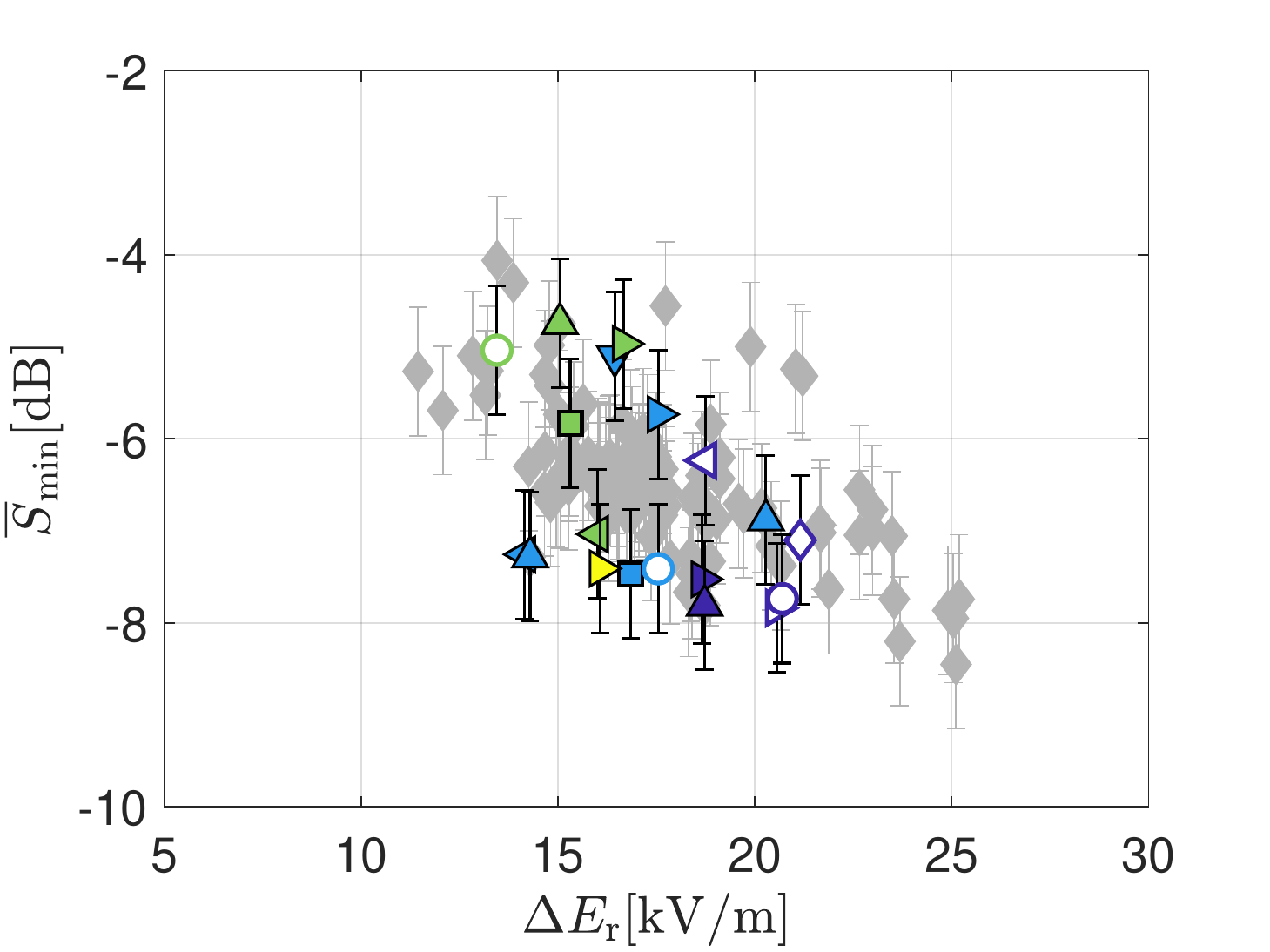}
\caption{$\overline{S}_{\rm min}$ against $\Delta E_{\rm r}$ for the DR scans given in tables \ref{table:dischargesDivertor} and \ref{table:dischargesSuppression}. The grey markers represent the extra DR scans that have been introduced in this study while the DR scans of the previous study are represented by the same markers as in figure \ref{figure:densityECHMaps}.}
\label{figure:turbulenceSuppression}
\end{figure}

As with $\overline{E}_{\rm r, SOL}$, we first study the evolution of $\overline{S}_{\rm min}$, calculated as explained in section \ref{section:experimental_setup}, with respect to basic plasma parameters, $\overline{n}_{\rm e}$, $P_{\rm ECH}$ and the ratio of the two quantities, $P_{\rm ECH} / \overline{n}_{\rm e}$. As can be seen in figure \ref{figure:SwithParameters}, there is no apparent trend between the $\overline{S}_{\rm min}$ and ECH power (plot \ref{subfigure:SwithECH}) nor between $\overline{S}_{\rm min}$ and the $P_{\rm ECH} / \overline{n}_{\rm e}$ ratio (plot \ref{subfigure:SwithECHDensity}). In figure \ref{figure:densityECHMaps}, the DR scans included in table \ref{table:dischargesSuppression} are depicted with grey markers while the DR scans included of table \ref{table:dischargesDivertor} using the same marker and color as in the figure above. In plot \ref{subfigure:Swithdensity}, some relation can be seen between $\overline{S}_{\rm{min}}$ and line averaged density, with an increment of about $5\rm{dB}$ over the observed density range (from $-9\rm{dB}$ to $-4\rm{dB}$ for an increment of the line averaged density from $4.5$ to $6.5 \times 10^{19} \rm{m^{-3}}$). This is particularly clear for the density ramps featured in the grey data. Finally, in figure \ref{figure:turbulenceSuppression}, $\overline{S}_{\rm min}$ is plotted against difference of the radial electric field at the sign reversal for the DR scans, $\Delta E_{\rm r}$, which is considered a proxy for the velocity shear at the edge. In it, an inverse relation between the two parameters is observed, arguably clearer than any of the ones seen in figure \ref{figure:SwithParameters}, indicating that there is some degree of turbulence suppression when the radial electric field shear becomes stronger. As well, the scattering of the points in the plot is found to decrease for increasing $\Delta E_{\rm r}$. However, it must be pointed out that this decrease is rather moderate: since $S \propto \delta n_{\rm e}^2$, the $5\rm{dB}$ decay observed over the whole range of shear values represents a reduction of less than a factor 2 in the density fluctuation amplitude. Interestingly, this factor is rather close to the ratio between the highest and lowest line averaged densities in the database. This would indicate that, if the relative amplitude of the fluctuations $\delta n_{\rm e}/n_{\rm e}$ does not change, the observed effect could be largely explained by the increase of density at the measurement points at the edge of the plasma (which roughly follows that of $\overline{n}_{\rm e}$).\\

\label{section:results}

\section{Discussion \& Summary}

The first objective of this work was to investigate the mechanism behind the formation of the electric field at the SOL of a stellarator with an island divertor. By analyzing divertor conditions over a set of standard configuration plasmas featuring a wide range of $\overline{E}_{\rm r, SOL}$ values in attached divertor conditions ($P_{\rm rad} / P_{\rm EC} < 0.5$), we have determined that the radial electric field at the SOL of the elliptic section follows the $\overline{E}_{\rm r, SOL} \propto T_e/\lambda_q$ relation expressed in equation \ref{eq:finalEr}. If some reasonable assumptions are made (namely, $\lambda_q \propto \lambda_T$ and small variation of the the incidence angle of the magnetic field at the target over the projection of the upstream measurement region, leading to $q_t \propto q_w$), this can be interpreted as a proof that parallel current and gradients of electron temperature and pressure are small and the upstream potential is determined by the sheath entrance potential under this conditions, $\phi_{u} \simeq \phi_{s}$. Since $\phi_{s} \simeq 3T_e$, this means that the potential profile in the SOL of W7-X in attached conditions is determined by the $T_e$ values at the divertor. This simple tokamak model is consistent with modelling carried out in the predecessor machine W7-AS \cite{Feng1999} and with previous observations at W7-X, which indicated that the main SOL drift for low density plasmas was caused by the radial electric field, $E_r \times B$ \cite{Hammond2020}. Therefore, our results imply that the intensity of the attached-SOL drifts would be dominated by the steepness of the $T_e$ profile at the target. Still, it must be taken into account that the DR used in this work can only measure at one position in the SOL of W7-X (corresponding to the bean-shaped plasma section shown in figure \ref{figure:methodPoincare}). Given the complex, nonaxisymmetric topology of the island divertor, it can not be assumed that this will be automatically the case for any other position and the determination of $\phi_u$ will be considerably more complicated for regions which are not directly connected to the targets. In order to obtain a full physical picture of the formation of the radial electric field, this study should be extended to plasmas with $P_{\rm rad} / P_{\rm EC} > 0.5$ (this is, with a lower  $P_{\rm ECH} / \overline{n}_{\rm e}$ ratio) featuring different degrees of detachment. If the relation to the proposed basic model holds, the link between $\overline{E}_{\rm r, SOL}$ and target conditions should disappear, and other factors (such as the local parallel currents)  should become dominant. This kind of investigation can not be carried out using the approach presented in the present study (and thus falls out of its scope), as the IR data can no longer be relied for it, and therefore the simple estimation of $\lambda_q$ used here is no longer possible. Instead, a complex combination of different diagnostics (such as Langmuir fixed probes at the targets, mid-plane manipulator, Helium beam, etc.) and simulations will be required to identify the SOL conditions which follow the variation of $\overline{E}_{\rm r, SOL}$ as measured by the DR in this higher density regime.\\ 

The second objective of this study was to determine whether the observed variations on the amplitude of the velocity shear observed across the LCFS have the expected suppression effect on the local turbulence. In this case, we combined measurements of the radial electric field and fluctuation amplitude profiles, both carried out by the DR. By doing so, we find that there is typically a depression of several dB in the profile of fluctuation amplitude which roughly coincides with the position of the change of sign of the radial electric field (where the shear would be at its maximum). This effect, which was shown for typical discharges in figure \ref{figure:exampleSuppression}, is indicative of the shear suppression mechanism. Then, we try to quantify the effect of the shear by comparing discharges with different density and heating power. As seen in figure \ref{figure:turbulenceSuppression}, we observe that fluctuation amplitude at the shear region do indeed decrease as the radial electric field jump between the SOL and the confined region becomes larger. However, this reduction in amplitude is rather moderate (the $5\rm{dB}$ observed in the figure correspond to a variation of around 80\% in $\delta n$) and just marginally above the uncertainty bars of the measurements. Moreover, given the already discussed evolution of the SOL electric field with plasma parameters, $E_{\rm r, SOL}$, and consequently $\Delta E_{\rm r}$, are reduced for discharges with higher density. Therefore, at least part of the apparent suppression of fluctuations can be simply explained by the reduction of local density at the measurement point (and indeed, fluctuation amplitude shows as well some correlation to line averaged density, as seen in figure \ref{figure:SwithParameters}). This result suggests that the effect on fluctuations of the velocity shear observed around the LCFS of W7-X is rather moderate, which is against expectations from studies at tokamaks, specially considering the non-negligible values of $\Delta E_{\rm r}$, which reach up to $25$ kV/m. In any case, the fact that fluctuation amplitude is clearly decreasing around the edge $E_{\rm r}$ shear for a given discharge (as seen for example in figure \ref{figure:exampleSuppression}), indicates that at least some degree of shear-related suppression is most likely taking place. Considering all of the above, it is difficult to achieve a strong conclusion here: Indeed, $\Delta E_{\rm r}$ is only a proxy for the shear which does not take into account the steepness of the $E_{\rm r}$ profile. The reason for this is the poor quality of density profiles at the edge of W7-X, which did not permit an accurate enough localization of the DR measurements in previous campaigns. This may be expected to improve in forthcoming campaigns thanks to upgrades on the TS system and the routine availability of alkali beam data \cite{Zoletnik2018}. While $\Delta E_{\rm r}$ allows for a qualitative discussion of the shear intensity (eg. between different discharges), it does not suffice for a comparison on the shearing rate with the autocorrelation time of the observed amplitude. As a consequence, it is difficult to know if the $k_{\perp}$ range of fluctuations observed by the DR in these discharges is below or above the spatial scale which should be decorrelated by the shear. As well, other examples exist of instances in which a strong radial gradient of $E_r$ seems to be followed by a strong local reduction of the fluctuation amplitude, as measured by the DR. One such case is the strong sheared structure caused in $u_{\perp}$ when an island chain of sufficient width is found inside the LCFS \cite{Estrada2021}. Therefore, while no clear evidence of shear-suppression effect on the turbulence has been found in this work, this analysis will need to be refined before it can be safely stated that there is no such mechanism at W7-X. In future experiments, upgrades for the DR system \cite{windischISHW2022} will be available in order to carry out a more detailed characterization of turbulence in the region, including wave number spectra and correlation studies which may shed light onto the effect of the shear on the elongation of eddies. Finally, beyond the specific effect of the shear on density fluctuations, its impact on turbulent transport should be systematically evaluated in order to determine its impact the overall performance of W7-X and thus assess the relevance of this phenomenon. \\

In summary, this study shows for the first time that the physics behind the formation of the electric field at the SOL of an island-divertor stellarator is qualitatively the same as in a tokamak, at least when the divertor is attached. This means that the radial electric field, which is responsible for the main SOL drifts in this regime, depends on the $T_{\rm e}$ values at the entrance of the sheath. The variation of the radial electric field around the separatrix of W7-X leads to a local reduction of the amplitude of density fluctuations, as expected from both models and experiments carried out in tokamaks. However, when different amplitudes of such shear -linked to the correspondingly different divertor conditions- are compared, a rather moderate effect on the fluctuations is observed. While this result initially hints that the edge shear could be a less efficient mechanism for the formation of edge transport barriers as it is in tokamaks, a more complete investigation is required before solid conclusions can be reached. This will be carried out in the forthcoming experimental campaigns of Wendelstein 7-X.\\

\label{section:conclusion}

\section{Acknowledgments}

The authors acknowledge the entire W7-X team for their support. This work has been partially funded by the Spanish Ministry of Science and Innovation under Contract No. FIS2017-88892-P and PID2021-125607NB-I00. This work has been sponsored in part by the Comunidad de Madrid under Project 2017-T1/AMB-5625. This work has been carried out within the framework of the EUROfusion Consortium, funded by the European Union via the Euratom Research and Training Programme (Grant Agreement No 101052200 - EUROfusion). Views and opinions expressed are however those of the author(s) only and do not necessarily reflect those of the European Union or the European Commission. Neither the European Union nor the European Commission can be held responsible for them.


\printbibliography

\end{document}